\documentclass[twocol]{ametsocV6.1}

\usepackage{amsmath}
\usepackage{bm}
\usepackage{algorithm}
\usepackage{algorithmic}
\usepackage{booktabs}
\usepackage[hidelinks]{hyperref}

\title{Efficient modeling of sub-kilometer surface wind with Gaussian processes and neural networks}

\authors{
    Francesco Zanetta\aff{a, b}\correspondingauthor{Francesco Zanetta, zanettaf@ethz.ch}
    Daniele Nerini\aff{b}
    Matteo Buzzi\aff{b}
    Henry Moss\aff{c}
}

\affiliation{
    \aff{a}{Institute for Atmospheric and Climate Science, ETHZ, Zürich} \\
    \aff{b}{Federal Institute of Climatology and Meteorology MeteoSwiss, Locarno-Monti} \\
    \aff{c}{Department of Applied Mathematics and Theoretical Physics, University of Cambridge, Cambridge} \\
}

\abstract{
Accurately representing surface weather at the sub-kilometer scale is crucial for optimal decision-making in a wide range of applications. This motivates the use of statistical techniques to provide accurate and calibrated probabilistic predictions at a lower cost compared to numerical simulations. Wind represents a particularly challenging variable to model due to its high spatial and temporal variability.
This paper presents a novel approach that integrates Gaussian processes and neural networks to model surface wind gusts at sub-kilometer resolution, leveraging multiple data sources, including numerical weather prediction models, topographical descriptors, and in-situ measurements.  Results demonstrate the added value of modeling the multivariate covariance structure of the variable of interest, as opposed to only applying a univariate probabilistic regression approach. Modeling the covariance enables the optimal integration of observed measurements from ground stations, which is shown to reduce the continuous ranked probability score compared to the baseline. Moreover, it allows the generation of realistic fields that are also marginally calibrated, aided by scalable techniques such as random Fourier features and pathwise conditioning. We discuss the effect of different modeling choices, as well as different degrees of approximation, and present our results for a case study.}

\begin{document}

\maketitle


\statement
Accurately representing the surface wind at the sub-kilometer scale is valuable for various applications, but it represents a difficult challenge, especially in the presence of complex topography. This study introduces a novel approach that combines machine learning approaches to model surface wind gusts, leveraging data from numerical weather prediction models, digital elevation models, and in-situ measurements. The methodology enables the generation of statistically optimal and realistic predictions of wind fields. Computationally efficient techniques are used and high prediction quality is achieved, making our proposed method a useful tool for high-resolution meteorological and climatological products. The method can be extended to other surface variables.


\section{Introduction}

Surface weather affects a wide range of human activities, from individual lives to critical infrastructure and operations. Among surface weather phenomena, the wind plays a crucial role, as it can be related to destructive weather conditions, such as dangerous storms, and affects important sectors such as aviation and energy production. Consequently, motivated by the need for optimal decision-making under uncertainty, there is a strong demand for accurate and precise information on the likely distribution of surface winds. However, surface wind phenomena, characterized by high spatial and temporal variability and the sparseness and limited representativeness of our measurement records, challenge our traditional modeling and monitoring techniques. As with precipitation in complex terrain \citep{lundquist_our_2019}, the path to more accurate surface wind estimates is through the coupling of multiple data sources and modeling methods.

Traditionally, applications that rely on surface wind analyses, that is, the accurate representation of surface winds on a dense spatial grid, have used numerical simulations from high-resolution atmospheric models \cite[e.g.,][]{bernhardt_using_2009, mott_meteorological_2010}. Numerical simulations, while central to surface wind analysis, face limitations due to their spatial and temporal discretization, leading to structural errors such as the misrepresentation of subgrid processes. Despite advances in improving resolution, physical parameterizations, and data assimilation, these models still struggle with systematic biases and inaccuracies \citep{vannitsem_statistical_2018,hemri_trends_2014}. In addition, improvements related to increased horizontal resolution typically require a significant increase in computational resources.

Statistical modeling stands as a complement to numerical simulations that has been used successfully in a wide range of applications in weather and climate. In the context of weather forecasting, a wide variety of techniques have been proposed to optimize the information from NWP models, many of which are classified as post-processing methods \citep[see][for a review]{vannitsem_statistical_2021}. These are typically used to statistically relate NWP model output and other covariates to observations, and obtain bias-corrected and calibrated predictions, with the advantage that they are very cheap to compute. In recent years, neural networks (NNs) have emerged as a valid option for this task, as they achieve state-of-the-art performance \citep{demaeyer_euppbench_2023} while providing great flexibility: they can accommodate diverse input formats (tabular data, temporal series, images, or sequences of images) and allow users to incorporate domain expertise when designing a model's architecture \citep[e.g.][]{zanetta_physics-constrained_2023, dujardin_windtopo_2022}.

One of the most important challenges when statistically optimizing NWP predictions is related to spatio-temporal consistency, that is, preserving the spatial and temporal correlation structure in post-processed predictions \citep{vannitsem_statistical_2021}. For the many standard approaches that predict univariate probability distributions (for a single variable, at a single location), where each prediction is treated independently, additional computationally involved steps \citep[e.g.][]{schefzik_uncertainty_2013, clark_schaake_2004} are required to reestablish correlation structures after sampling. The second crucial challenge is the question of how to blend diverse sources of predictability, such as NWP model outputs and observational data. A key intuition is that both of these challenges are fundamentally related to spatio-temporal covariance modeling, thus making geostatistical methods an attractive option.

A well-established technique traditionally employed in geostatistical modeling is Kriging. It is designed specifically for the purpose of interpolating spatial data. In essence, Kriging makes the assumption that spatial variation in the data can be captured by a statistical model known as a variogram, which quantifies the degree of spatial correlation between data points based on their separation distance. In the context of machine learning (ML), Kriging is frequently referred to as Gaussian processes (GPs) regression, which can sometimes lead to confusion for researchers transitioning between fields. However, there are notable distinctions between Kriging and GPs. GPs offer more flexible covariance modeling via the use of kernels, expanding beyond the traditional variogram models in Kriging, which, combined with optimization using maximum likelihood methods, can lead to potentially more accurate model predictions  \citep[e.g.][]{christianson_traditional_2023}. Moreover, while Kriging is typically implemented in two or three spatial dimensions, GPs shine in their ability to generalize to higher-dimensional and potentially non-euclidean spaces. In summary, while both Kriging and GPs focus on modeling covariance between points, the latter presents a broader and more versatile framework, enriched by the vast ecosystem of ML \citep[e.g.][]{gardner_gpytorch_2021, matthews_gpflow_2017, pinder_gpjax_2022}. 

In this work, we combine ideas from NN-based statistical post-processing with GPs. We introduce a novel methodology integrating GPs and NNs that uses NWP data with additional covariates, such as high-resolution topographical information and observed wind gust measurements at sparse locations. Recognizing the computational demands of traditional GPs, especially when inference is performed on large datasets such as dense grids, our approach implements scalable GP techniques. In particular, we take advantage of random Fourier features \citep[RFF]{rahimi_random_2007}, an approximation method that significantly reduces the computational overhead of GPs. Furthermore, we employ pathwise conditioning from \cite{wilson_pathwise_2021}, which facilitates efficient posterior sampling from the Bayesian model. We discuss our results both qualitatively, presenting realizations of wind fields on a grid for a specific case study, and quantitatively, by evaluating our models on an independent test set using standard verification metrics and diagnostic tools.

\section{Data and methods}

\subsection{Data}
This study uses three different types of data sets: NWP model data, digital elevation model (DEM) data, and observational data, all spanning January 2020 to December 2023.
\begin{itemize}
    \item \textbf{NWP}: Our NWP data comes from COSMO-1E hourly analysis --  a regional forecasting model that is used operationally in MeteoSwiss. In its high-resolution probabilistic configuration, COSMO-1E runs a 11 member ensemble with a horizontal resolution of 1.1 km and an update cycle of 3 hours. The hourly analysis is obtained via the kilometer-scale ensemble data assimilation system \citep[KENDA,][]{schraff_kilometrescale_2016}, which implements an assimilation scheme based on the ensemble Kalman filter. Note that while some surface wind measurements are assimilated by KENDA, none fall within our study domain. For our use case, we only consider the analysis from the control run. NWP data are bilinearly interpolated on target points.
    \item \textbf{DEM}: Our DEM has a horizontal resolution of 50 meters, from which we derived a set of topographic descriptors known to be related to the dynamics of surface wind, such as the topographical position index, the maximum upwind slope \citep{winstral_spatial_2002} or directional (NS and EW) gradients. Data are interpolated bilinearly at the target points.
    \item \textbf{Observations}: Observational data is obtained from a network of more than 500 meteorological stations coming from different sources. These stations are distributed throughout central Europe, with the majority located in Switzerland (see Fig. ~\ref{fig:partitioning}), and achieve a broad coverage of the region's diverse weather patterns and local conditions. These observational data comprise hourly wind gust measurements, typically defined as the 1\,second average wind maximum. To ensure the consistency and reliability of our results, the observational dataset was carefully quality-controlled. We excluded all stations with anemometers below 6 or above 15 meters height above the ground, as well as obvious measurement errors such as negative wind speeds or plateaus of constant values. In addition, a manual validation for suspicious measurements was conducted by cross-checking measurements from adjacent stations, providing a system to identify and correct anomalies.
\end{itemize}

\subsection{Neural network-based postprocessing baseline}
As a baseline for our results presented in Section~\ref{sec:results} we use a simple NN-based post-processing method (NNPP), similar to the approach presented in \citet{rasp_neural_2018}. Here, a NN is trained to predict the marginal distribution at each prediction point, using the continuous ranked probability score (CRPS) as a loss function. For predictors, the NWP direct model output is used along with other covariates such as topographical descriptors and temporal information (sine and cosine components of the hour of the day). While several ways exist to implement a probabilistic regression for this kind of application, we have chosen one that is as close as possible to the methods presented in the next section, in order to make them more comparable. Therefore, we performed the regression on the transformed data (see Appendix~\ref{appendix:transformation}) using the CRPS for the normal distribution as loss function.

\subsection{Surface weather analysis with Gaussian Processes regression}

\newcommand{\bx}{\boldsymbol{x}}
\newcommand{\bX}{\boldsymbol{X}}

Let us start by introducing the GP approach to regression, which will be connected later to our problem. Consider a regression task $y = f(\bx) + \epsilon$ where $y \in \mathbb{R}$ is the target variable, $x \in \mathbb{R}^D$ the inputs, and $\epsilon \sim \mathcal{N}(0, \sigma_{\epsilon}^2)$ some additional independent Gaussian observational noise. We assume that $f(\bx)$ can be described as a Gaussian process, i.e. any finite collection of $f(\cdot)$ follows a particular multivariate normal distribution. The process is then entirely defined by a mean function $\bm{\mu_x} = m(\bx)$ parameterized by $\boldsymbol{\theta}_m$ and a covariance kernel $\bm{K_{xx'}} = k(\bx, \bx')$ parameterized by $\boldsymbol{\theta}_k$ 
\begin{equation} \label{eq:gpprior}
    f(\cdot) \sim \mathcal{GP}(m(\cdot), k(\cdot, \cdot')).
\end{equation}
The mean function and the covariance kernel incorporate our knowledge of the data and can be used as a prior distribution from a Bayesian inference perspective. Given a set of $n$ observed locations (or \textit{context} points) $ \mathcal{D} = (\bX, \boldsymbol{y}) = \{\bx_i, y_i\}_{i=1}^n$, we can leverage Bayes' theorem to update $\mathcal{GP}$ and obtain the posterior distribution conditioned on $\mathcal{D}$, effectively correcting the mean function and reducing the uncertainty of our predictive distribution.

To calculate our GP posterior, we have to consider the resulting joint distribution between the observed locations and the function values at a new target locations $\bm{f_*}=f(\bX_*)$:

\begin{equation}
    \begin{bmatrix}
        \bm{y} \\ \bm{f_*} 
    \end{bmatrix} 
    \sim 
    \mathcal{N} \left( 
    \begin{bmatrix}
        \bm{\mu_x} \\ \bm{\mu_*}
    \end{bmatrix},
    \begin{bmatrix}
         \bm{K_{xx}} + \sigma_{\epsilon}^2I & \bm{K_{x*}} \\ \bm{K_{*x}} & \bm{K_{**}}
    \end{bmatrix}
    \right)
\end{equation}

where $\bm{\mu_x} = m(\bm{X})$ and $\bm{\mu_*} = m(\bm{X_*})$ are the mean function values for the observed and target points, respectively, $\bm{K_{xx}}=k(\bX, \bX)$ is the covariance matrix, $\sigma_{\epsilon}^2$ is the noise variance, and $\bm{K_{x*}} = k(\bX, \bX_*)$ represents the cross-covariance between context and target points. When the joint distribution is a multivariate Gaussian, the marginalization and conditioning with respect to $\mathbf{y}$ is exact, and the predictive distribution of $f_{*|\mathbf{y}}$ for target points remains Gaussian and has mean and covariance:
\begin{equation} \label{eq:gpupdate}
    \begin{aligned}
    & \mu_{*|\bm{y}} = \bm{\mu_*} + \bm{K_{*x}}[\bm{K_{xx}} + \sigma_{\epsilon}^2I]^{-1}(\bm{y} - \bm{\mu_x}), \\
    & \Sigma_{*|\bm{y}} = \bm{K_{**}} - \bm{K_{*x}}[\bm{K_{xx}} + \sigma_n^2I]^{-1}\bm{K_{x*}}.
    \end{aligned}
\end{equation}
Furthermore, the log marginal likelihood can also be expressed analytically and is given by
\begin{equation} \label{eq:mll}
    \begin{aligned}
    \log p(\bm{y} | \bX) = & -\frac{1}{2}(\bm{y - \mu_x})^T [\bm{K_{xx}} 
    + \sigma_{\epsilon}^2I]^{-1}(\bm{y - \mu_x})  \\
    & - \frac{1}{2}\log |\bm{K_{xx}} + \sigma_{\epsilon}^2I| \\
    & - \frac{n}{2}\log 2\pi,
    \end{aligned}
\end{equation}
which is used as a loss function during the optimization of a model's parameters. Let $\boldsymbol{\xi} = \{\boldsymbol{\theta}_{\mu}, \boldsymbol{\theta}_{k}, \sigma^2_{\epsilon}\}$ be the set of parameters of our GP, then we use gradient-based optimization on the following objective
\begin{equation}
    \boldsymbol{\xi}^{\star} = \operatorname{argmin}_{\boldsymbol{\xi} \in \Xi} - \log p(\mathbf{y})\,.
\end{equation}
Now that we have introduced the theoretical background, let us consider our application. First, we aim to train a model to use NWP data, topographical and temporal information as inputs $\bx$ to predict a surface weather variable $y$ (in our case wind gust), using station measurements as ground truth. Moreover, we are interested in modeling several points $X = {x_1, x_2, ..., x_n}$ and their joint probability distribution instead of each target location independently, which requires a covariance function $k(\bm{x, x'})$. Combined with a mean function $m(\bm{x})$, we have model for our prior probability distribution of $f$ as seen in Eq.~\ref{eq:gpprior}. Note that while the GP model as a whole takes all the input features $\bm{x}$, different features are then selected for the mean function or the covariance kernel. For instance, geographical coordinates are part of the input, but are only used by the kernel.

The next step is then to update the prior probability distribution of $f$ with the points observed from the ground stations, represented by $\mathcal{D}$, applying Eq.~\ref{eq:gpupdate}. This allows us to obtain probabilistic forecasts that are conditioned on surface measurements. The intuition is quite simple: prior forecasts at target locations will receive an update that is proportional to the covariance with the observed locations and inversely proportional to $\sigma^2_{\epsilon}$. We note that this update can also be viewed as an interpolation of the residuals $y -\mu_x$ from the mean function of our prior, and those familiar with geostatistical methods may see that this is conceptually close to regression Kriging.
We now proceed to briefly discuss the individual components of our methodology, namely the mean function, the covariance function, and observational noise.

\subsubsection{Mean function}
The function $\mu: \mathbb{R}^D \rightarrow \mathbb{R}$ calculates the average of $f$ for the $D$-dimensional input $\bx$. In our application, we show that this function can be implemented as a neural network parameterized by the weights $\boldsymbol{\theta_{\mu}} = \mathbf{w}$. When the NWP forecast is part of $\bx$, the mean function can be seen as a bias correction step applied to the NWP forecast. This correction is learned based on information from topographical descriptors and temporal encodings, as well as their interplay. For example, these features indicate whether the target location is sheltered or exposed to the incoming wind. To highlight the added value of the NN-based mean function with respect to simpler alternatives, we present results for multiple configurations in Section~\ref{sec:results}. There are two approaches to optimize $\boldsymbol{\theta_{\mu}}$: the first involves training these parameters concurrently with $\boldsymbol{\theta_{k}}$ and $\sigma^2_{\epsilon}$, while the second approach involves pretraining the mean function separately and optionally keeping the parameters fixed during the optimization of the posterior GP. We have found that pretraining the mean function generally leads to better results.

\subsubsection{Covariance function}
The GP framework provides a highly flexible approach to model covariance between points through a variety of kernel functions, both stationary and non-stationary. In stationary kernels the covariance only depends on the shift between two input values and not the values themselves. A common example of such type of kernels is the radial basis function (RBF) kernel
\begin{equation}
    k(\bx, \bx') = \sigma^2 \exp\left(-\frac{||\bx - \bx'||^2}{2\boldsymbol{l}^2}\right),
\end{equation}
where $l$ and $\sigma^2$ are the length-scale and variance parameters $\boldsymbol{\theta}$ of the kernel. On the other hand, non-stationary kernels allow the output covariance to depend on the input values themselves, and a prominent example is the linear kernel
\begin{equation}
k(\bx, \bx') =  \sigma^2 \bx \bx^T.
\end{equation}
Kernels can even be combined with NNs by learning non-linear transformations of the input space that map to a latent space where a kernel is applied \citep{wilson_deep_2015}.
Starting from a base kernel $k\left(\mathbf{x}_i, \mathbf{x}_j \mid \boldsymbol{\theta_{base}}\right)$ with parameters $\boldsymbol{\theta_{base}}$ the input $\bx$ can be transformed as
\begin{equation} \label{eq:deepkernel}
    k\left(\bx, \bx' \mid \boldsymbol{\theta_{base}}\right) \rightarrow k\left(g\left(\bx, \mathbf{w}\right), g\left(\bx', \mathbf{w}\right) \mid \boldsymbol{\theta_{base}}, \mathbf{w}\right),   
\end{equation}
where $g(\bx, \mathbf{w})$ is a non-linear mapping given by a NN parameterized by weights $\mathbf{w}$.
Furthermore, multiple kernels can be combined into more expressive ones to account for additive or multiplicative effects of the different factors contributing to the modeled process. Indeed, the sum of two kernels is a kernel, $k(\bx, \bx') = k_1(\bx, \bx') + k_2(\bx, \bx')$, and a product of two kernels is a kernel,  $k(\bx, \bx') = k_1(\bx, \bx')k_2(\bx, \bx')$ \citep{rasmussen_gaussian_2006}. 

For the case of wind gusts and spatially distributed phenomena in general, a good first assumption is that the covariance simply depends on distance, and one might consider a kernel with spatial coordinates and elevation as input. However, there are cases where the distance alone cannot fully describe the actual covariance between two locations: it might also depend by their specific geomorphological setting or on dynamic factors such as the direction or intensity of the wind, and their interplay with the diurnal cycle. This is where the flexibility of the kernel construction we have just described comes to help: we can design much more complex kernels, capable of including a diverse set of input variables. In Section~\ref{sec:results} we present three variants of kernel functions.

\subsubsection{Observational noise}
As the name suggests, the observational noise represents the uncertainty of the observed measurement $y$. It is analogous to the \textit{nugget} parameter in the context of Kriging. When accounted for, it can make the GP model more robust to outliers and noise in the data. This is because the noise term in the GP model acknowledges that the observed data might not be perfectly accurate. Furthermore, the presence of noise influences the model's predictive uncertainty. In areas where the model has observed (noisy) data $y$, it will still have some uncertainty, and the function $f_{|y}$ will not be exactly $y$. This also has an effect on the "smoothness" of the predictions: by accounting for some noise, $f_{|y}$ will vary more smoothly around the observed data.
Importantly, observational noise requires scaling due to the non-linear transformation performed on the target data distribution (to make it standard normal), as explained in Appendix~\ref{appendix:transformation}. The observational noise parameter can either be learned during optimization or be a fixed value arbitrarily chosen by the user based on the importance they give to observed data, which should reflect the uncertainty of the measuring instruments. In our experiments, we have observed that keeping the observational noise constant led to a more stable training. For this reason, we have chosen to fix the observational noise $\sigma_{\epsilon}^2$ to 1\,$ms^{-1}$ (in the original, untransformed space). Using an observational noise that is explicitely derived from an empirical determination of measurement uncertainty would be more ideal, but we preferred to opt for an educated estimation for ease of implementation.

\subsection{Problem definition, training and evaluation setup}
In defining the problem, we start with the simplifying assumption that data from different timesteps $t$ are independent. In other words, we only focus on modeling spatial dependencies and do not explicitly consider temporal ones. In a single time step $t$, our data consists of $n$ spatially distributed input points $\{\bx_1, \bx_2, ..., \bx_n\} \in \mathbb{R}^D$ and corresponding outputs $\{y_1, y_2, ..., y_n\} \in \mathbb{R}$. For simplicity, we denote them as $\Vec{X} \in \mathbb{R}^{n \times D}$  and $\Vec{Y} \in \mathbb{R}^{n}$, and we denote $\mathcal{D}_t = (\Vec{X}_t, \Vec{Y}_t)$ as the set of points at time $t$. A single timestep therefore represents a single GP regression task: we are not interested in modeling the covariance across different timesteps but only the spatial covariance. 

Before we discuss our training strategy, let us describe the data partitioning method. Our dataset is split into training, validation, and test sets. We used 70\% of the stations and 2 years of data for training, 10\% of stations and 1 year for validation and 20\% and 1 year for testing. For the validation and test sets, we want to evaluate the performance of the model at a set $\mathcal{T}$ of \textit{target} stations given a set $\mathcal{C}$ of observed \textit{context} stations. Therefore, we represent an evaluation task as $\mathcal{D}^{eval} = \{(\Vec{X}^t, \Vec{Y} ^t), (\Vec{X}^{c}, \Vec{Y}^{c})\} = \{\mathcal{T}, \mathcal{C}\}$. Note that context stations are points that the model has seen during training, which are used to predict on new locations (targets) during evaluation. Further details on the partitioning method are provided in Appendix~\ref{appendix:partitioning}. After splitting our dataset, we standardize our input using the mean and standard deviation of the training set, and more importantly, we also transform the output values to a standard normal distribution (see Appendix~\ref{appendix:transformation} for details). 

In terms of training strategy, a naïve approach would be to fit a different GP model to each individual timestep. However, it has the inconvenience that the optimization procedure must be repeated several times independently -- which is a computational burden, leads to varying results, and is potentially more prone to overfitting due to the small number of points. A more robust approach is to optimize a model that achieves good performance across many tasks. This can be done by considering a metalearning approach, where the training loop is applied on tasks. Our method is similar to the one described by \citet{patacchiola_bayesian_2020}, with the difference that at each step we update the model parameters using mini-batches of tasks instead of a single task. The pseudocode is given in Algorithm~\ref{alg:training}. Let $\boldsymbol{\mathcal{D}} = \{\mathcal{D}_t\}_{t=1}^T$ be the complete training data set consisting of $T$ tasks and $\boldsymbol{\xi}$ the set of model parameters. At each training step, we sample a batch of tasks $\mathcal{D}_i \sim \boldsymbol{\mathcal{D}}$, and for all tasks we compute the negative log marginal likelihood using Eq.~\ref{eq:mll}. We then take the average of the loss for the entire batch and use it to perform the parameters update based on gradient descent. During training, we also computed the loss in a validation set. More details about the training procedure are presented in Appendix~\ref{appendix:training}.

\newcommand{\bxi}{\boldsymbol{\xi}}
\begin{algorithm} 
\caption{Optimizing GP model parameters across all tasks}
\begin{algorithmic}[1] \label{alg:training}
\REQUIRE train dataset $\boldsymbol{\mathcal{D}} = \{\mathcal{D}_t\}_{t=1}^T$
\REQUIRE model parameters $\boldsymbol{\xi}$
\STATE initialize model parameters
\WHILE{not done}
    \STATE Sample batch of $N$ tasks $\mathcal{D}_i \sim \boldsymbol{\mathcal{D}}$
    \FORALL{$\mathcal{D}_i$} 
        \STATE Compute $\mathcal{L}_i = -\log p(\mathcal{D}_i,\bxi)$
    \ENDFOR
    \STATE Compute $\mathcal{L} = \frac{1}{N} \sum_{i = 1}^N \mathcal{L}_i$
    \STATE Update $\bxi \leftarrow \bxi - \alpha \nabla_{\bxi}\mathcal{L}$
\ENDWHILE
\end{algorithmic}
\end{algorithm}

\subsection{Scalability to large grids}

A critical aspect concerning GPs is that they typically do not scale well with the amount of data. Specifically, the exact computation requirements of the GP algorithm scale cubically in the number of training/context points, and storage requirements (for the covariance matrix) follow a quadratic increase. Fortunately, recent work has provided a variety of new approximation methods that greatly improve the computational efficiency of GPs \citep{liu_when_2020}.

In addition to the poor scaling of exact GPs in the number of training/context points, similar limitations are found when building realizations of the model across large sets of target points. In the context of spatiotemporal modeling this can become quite challenging as predictions are often made on a dense grid. To grasp the scale of the numbers involved, consider a target grid of $1000\times1000$ pixels, with $n=1'000'000$ input points $\bX = \{\bx_1, \bx_2, ..., \bx_n\}$. To obtain the full multivariate distribution, one would need to compute a covariance matrix $K(\bX, \bX)$ of size $n^2$. When working with double precision floats, which is often the case in the context of GPs due to numerical stability, this would require to store an 8-\textit{terabyte} large matrix. For all practical purposes, this is intractable.

\subsubsection{Random Fourier features} \label{sec:rff}

A viable solution to the scalability problem that we have just presented is to avoid computing the full covariance matrix altogether via random Fourier features \citep[RFF]{rahimi_random_2007} methods. This technique allows for the approximation of shift-invariant kernel functions, making it especially advantageous for GPs employing commonly used kernels such as the RBF or the Matern family kernels. Given a shift-invariant kernel $ k(\mathbf{x}, \mathbf{x'}) = k(||\mathbf{x} - \mathbf{x'}||)$ where $ \mathbf{x} $ and $ \mathbf{x'} $ are input vectors, RFF provides an approximation
\begin{equation} \label{eq:rffinner}
     k(\mathbf{x}, \mathbf{x'}) \approx \mathbf{\phi}(\mathbf{x})^\top \mathbf{\phi}(\mathbf{x'}).   
\end{equation}
The kernel can be understood as an inner product in an infinite-dimensional feature space approximated by a finite number of Fourier components $L$, and $\boldsymbol{\phi}: \mathcal{X} \rightarrow \mathbb{R}^{L}$ is the feature map to the $L$-dimensional approximation of the said feature space. The feature map is defined as

\begin{equation} \label{eq:featuremap}
  \mathbf{\phi}(\textbf{x}) := \sqrt{\frac{2}{L}} \left[\begin{array}{c}
\sin \left(\boldsymbol{\omega}_1^{\top} \mathbf{x}\right) \\
\cos \left(\boldsymbol{\omega}_1^{\top} \mathbf{x}\right) \\
\vdots \\
\sin \left(\boldsymbol{\omega}_{L / 2}^{\top} \mathbf{x}\right) \\
\cos \left(\boldsymbol{\omega}_{L / 2}^{\top} \mathbf{x}\right)
\end{array}\right], \quad \boldsymbol{\omega}_i \stackrel{iid}{\sim} P(\omega),  
\end{equation}

with $P(\omega)$ denoting the spectral density of the kernel (interpreted as a probability distribution over frequencies) and $L$ the number of random features. For instance, the spectral density $P(\omega)$ of the RBF kernel is the standard Gaussian distribution, and each $\omega_i$ is an $i.i.d.$ sample from said distribution.
Then, it is possible to define a random function of a GP whose covariance is approximately $k(\cdot, \cdot)$

\begin{equation} \label{eq:fnsample} 
    f(\cdot) \approx \mu(\cdot) + \boldsymbol{\phi}(\cdot)^{\top} \mathbf{w}=\mu(\cdot)+ \sum_{i=1}^{L} w_i \phi_i(\cdot), \quad w_i \sim \mathcal{N}(0, 1)\,.
\end{equation}

Importantly, this formulation allows us to draw realizations from an approximate GP prior in linear complexity with the number of inference points, without the need to compute the covariance matrix. In practice, we have a function that can be realized anywhere in our input space, where the stochasticity is entirely controlled by $\mathbf{w}$. This also means that during inference the computation of a realization can be divided into chunks, to further reduce memory usage, and since $\mathbf{w}$ is kept constant, the aggregated chunks will still yield a coherent sample, without any artifact at the chunk boundaries.

As shown in the previous section, it is possible to combine multiple kernels via summation or multiplication. In our study we consider the product combination for our kernels, and this can also be implemented when using the RFF approximation. Consider two shift-invariant kernels $k_1$ and $k_2$. Their respective feature maps are defined following Equation~\ref{eq:featuremap}. We obtain the product of the two kernels by combining the sampled frequencies $\Vec{\omega}_1$ and $\Vec{\omega}_2$ as
\begin{equation}
\Vec{\omega}_{\text{combined}} = \sqrt{\Vec{\omega}_1^2 + \Vec{\omega}_2^2},
\end{equation}

and subsequently apply Equation~\ref{eq:featuremap}.

\subsubsection{Pathwise conditioning}
Gaussian posteriors are typically obtained via a location-scale transform applied on mean vectors and covariance matrices of the prior distribution. The avenue for prior sampling offered by RFF allows us to express Gaussian posteriors in a different but equally valid way, which focuses on samples (or \textit{paths}) instead of distributions. Such an approach is presented in \citet{wilson_pathwise_2021}. To compute a posterior sample $(f\mid\boldsymbol{y})(\cdot)$, a two steps procedure is required: first, a prior sample at the target location $f(\cdot)$ is generated, and the same sample is also evaluated at the observed location $f(X)$. Then, an update term is computed based on the difference between the $f(X)$ and the observed value, and it is then added to the prior to yield the posterior sample conditioned on the observed data. This procedure is described by

\begin{equation}
    \underbrace{(f \mid \boldsymbol{y})(\cdot)}_{\text {conditional }} \stackrel{\mathrm{d}}{=} \underbrace{f(\cdot)}_{\text {prior }}+\underbrace{K\left(\cdot, \mathbf{X}\right) K(\mathbf{X}, \mathbf{X})^{-1}\left(\boldsymbol{y}-f(\mathbf{X})\right)}_{\text {update }},
\end{equation}

approximated by 

\begin{equation}
    (f\mid\boldsymbol{y})(\cdot) \approx \sum_{i=1}^L w_i \phi_i(\cdot) + \sum_{j=1}^n v_j k(\cdot, \boldsymbol{x}_j),
\end{equation}

where $\boldsymbol{v} = \mathbf{K}^{-1}_{n,n}(\boldsymbol{y} - \mathbf{\Phi}\boldsymbol{w})$. The great advantage of this way of obtaining posterior samples is that it scales linearly in the number of target points while allowing splitting computation into chunks.

\newcommand{\bs}{\boldsymbol{x}}

\begin{figure*}[t]
\includegraphics[width=\textwidth]{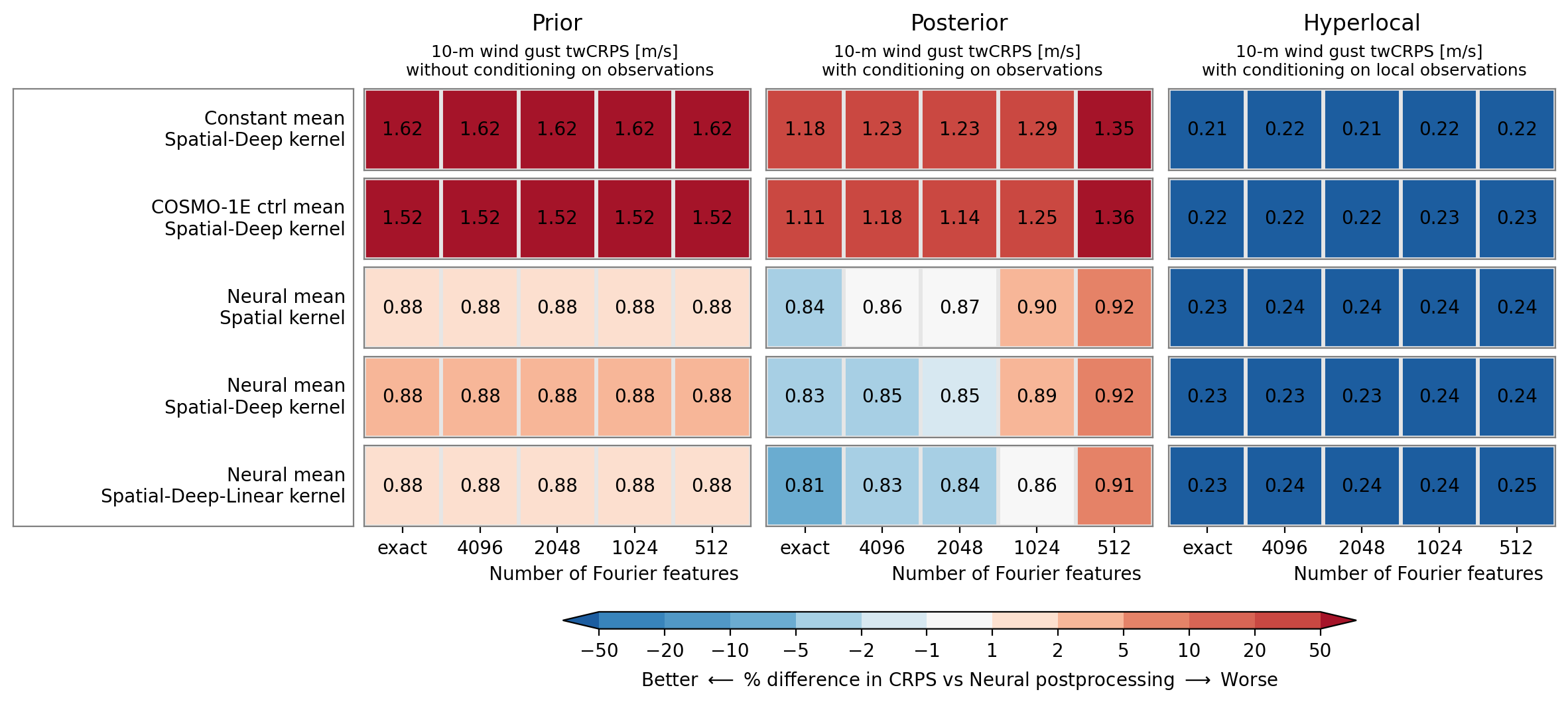}
\caption{Scorecard summarizing the evaluation metrics on the test dataset. Values in each box indicate the thresholded (at 4\,m/s) CRPS, and colors indicate the relative performance with respect to the baseline, as shown in the colorbar.}
\label{fig:scorecard}
\end{figure*}

\section{Experiments and results}
\label{sec:results}

We have designed an experimental setup with the goal of presenting results for different modeling choices, degrees of approximation, and type of predictions.

Differences in modeling choice concern how the mean function and the covariance function are defined. This can be highly flexible, allowing users to adapt their model's complexity or to account for domain knowledge instead of purely relying on data-driven optimization. Although a large variety of solutions is possible, we present here only a few combinations that allow us to draw some general conclusions.
By degrees of approximation we mean the number of Fourier features used in the RFF approximation of the covariance kernel.
The type of prediction is a distinction between using the prior predictive distribution (without conditioning on the observed data), the posterior predictive distribution (with conditioning on context, nearby stations) and the \textit{hyperlocal} predictive distribution, which is the prediction at locations where the observation itself is available and used.

These results, all based on the test dataset $\mathcal{T}_{test}$, are summarized with a scorecard in Fig.~\ref{fig:scorecard} in terms of the threshold-weighted CRPS (twCRPS, with threshold at 4 $ms^{-1}$) and its relative skill, color-coded, compared to the plain NNPP approach chosen as a baseline. For the NNPP approach we have a twCRPS of 0.87\,$ms^{-1}$ Additionally, we evaluate the calibration of our models by the aid of conditional probability integral transform \citep[cPIT]{allen_weighted_2022} histograms and reliability diagrams (for the probability of exceeding thresholds of 10\,$ms^{-1}$, 20\,$ms^{-1}$ and 30\,$ms^{-1}$). More details on the evaluation metric and diagnostic tools can be found in Appendix~\ref{appendix:evaluation}.

\subsection{Mean modeling}
\label{sec:mfexp}

\begin{figure*}[t]
\includegraphics[width=\textwidth]{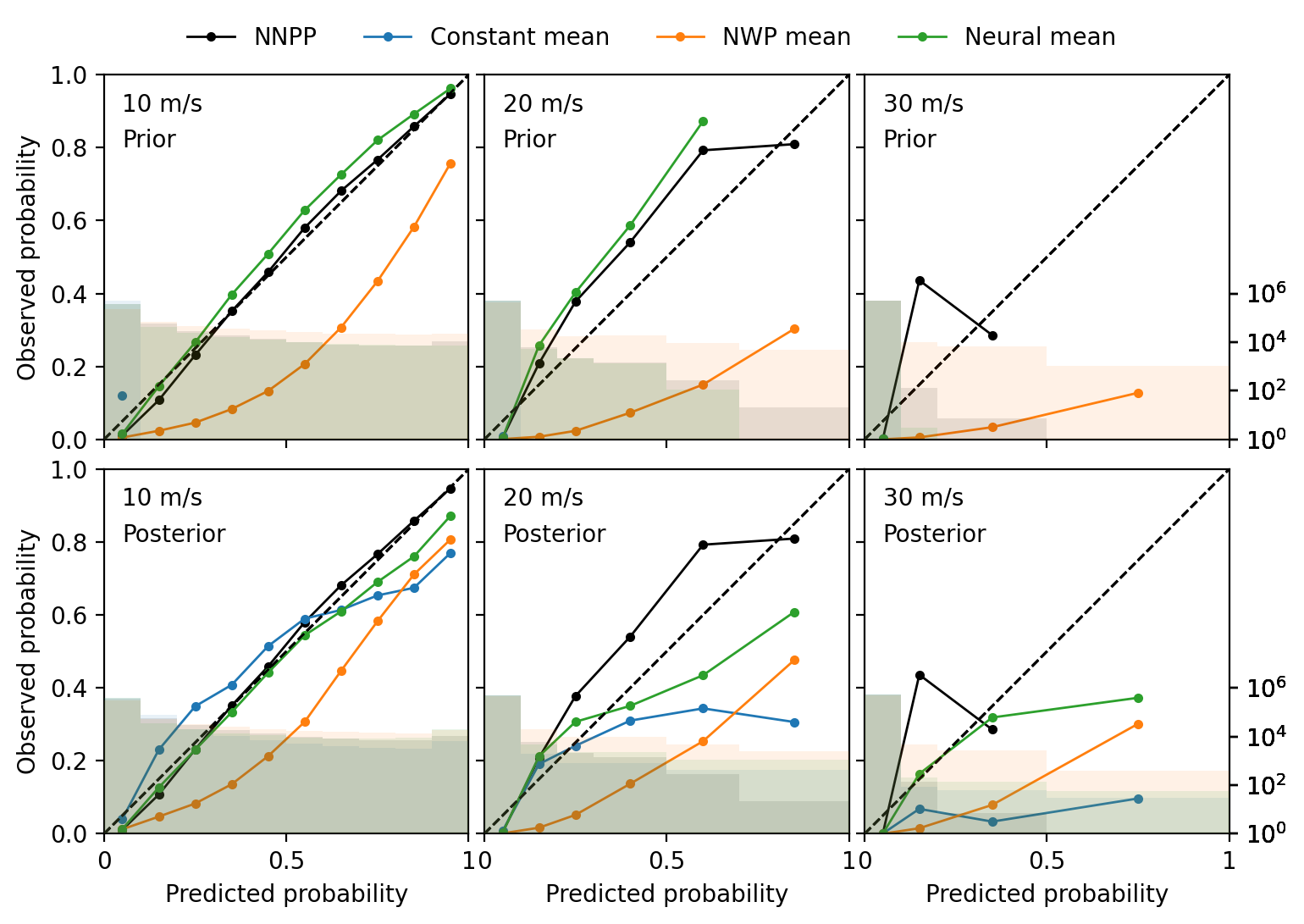}
\caption{Reliability diagrams for the mean modeling experiments, with columns representing the different exceedance thresholds (10, 20 and 30\,$ms^{-1}$) and the first and second rows are for prior and posterior predictions respectively. The secondary axis shows the log-scaled number of occurrences for each bin.}
\label{fig:mean_exp_rel}
\end{figure*}

\begin{figure*}[t]
\includegraphics[width=\textwidth]{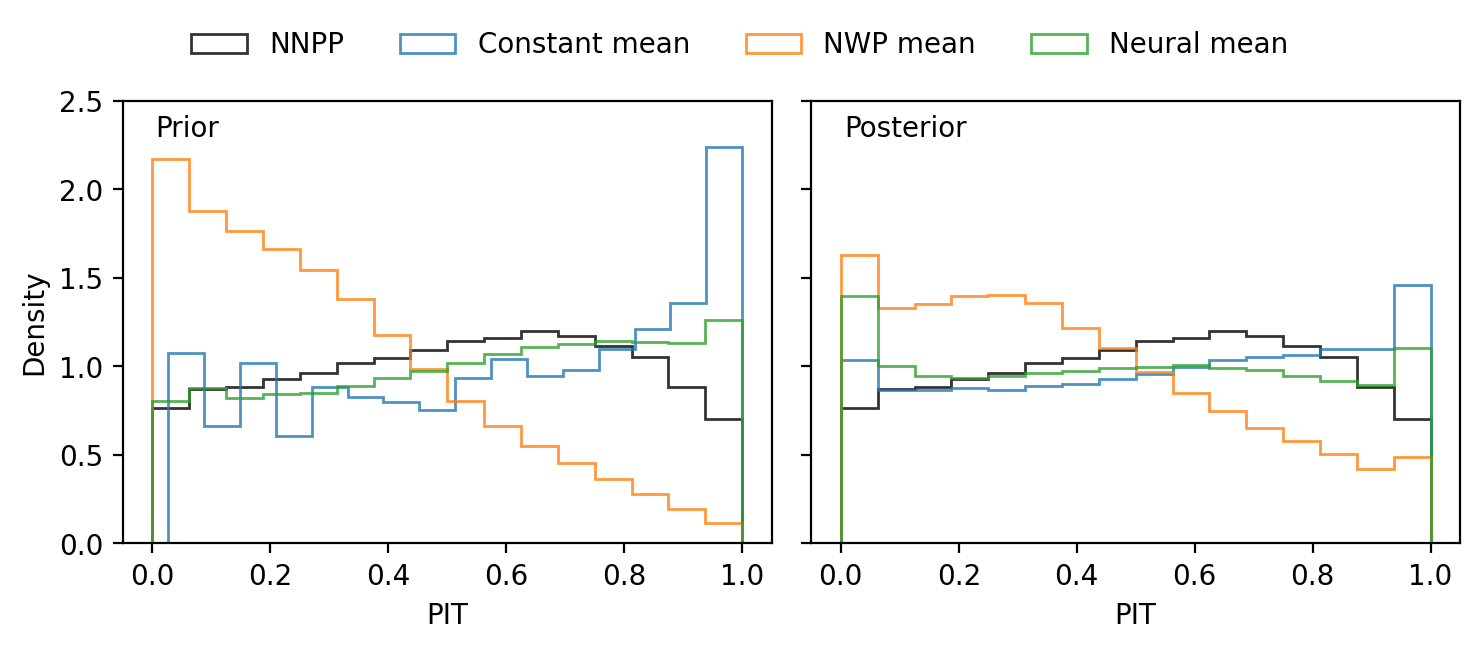}
\caption{cPIT histograms for the mean modeling experiments, with the results for the prior and posterior predictions on the left and right hand side respectively. Perfect calibration would be indicated by a uniform histogram.}
\label{fig:mean_exp_pit}
\end{figure*}

For the mean function, we have considered three approaches in which increasingly more information is added to the prior: a learned constant value, a mean function based on the raw COSMO-1E wind gust, and a mean function based on the bias corrected (via a NN) COSMO-1E wind gust. In all cases, the same kernel function $k_{SD}$, described below, was used. It is our purpose here to demonstrate the effect of including more prior information whenever it is available, since it is common in the GP literature to prioritize the modeling of the covariance rather than the mean function of the prior. 
In Fig.~\ref{fig:scorecard}, the considered approaches are on the first, second and fourth row respectively. Focusing on the exact GP, without RFF approximation, we observe that the neural mean approach significantly outperforms the constant mean and NWP mean approaches, and is the only approach that, for posterior predictions, shows positive skill compared to the NNPP baseline.

Figure~\ref{fig:mean_exp_rel} shows the reliability diagrams for the models considered in this experiment. In the top row, the diagrams for the prior predictions show two interesting aspects: first, the constant mean approach performs poorly, as it barely predicts any probability above 10\%. This is to be expected because the model simply learned a global climatology and does not exhibit any sharpness, therefore missing out large values completely. The NWP mean approach, on the other hand, overpredicts wind gust for all thresholds. This is consistent with the error structure of COSMO-1E itself, which tends to overestimate the wind gust, especially in complex topography. Prior predictions for the neural mean approach are well calibrated for the 10\,m/s threshold, but underestimate probabilities for the other thresholds. Figure~\ref{fig:mean_exp_pit} shows the PIT histogram for the same experiment, and the results for the prior, on the left-hand side, confirm what we have observed in the reliability diagrams: the constant mean approach completely misses large values, as indicated by the increased density in the right-most bin; the NWP mean approach has a positive bias, as indicated by the slope of the PIT histogram; the neural mean approach has the best overall calibration, but still shows a slight negative bias.

When considering the posterior predictions, as shown in the bottom row of  Fig.~\ref{fig:mean_exp_rel}, we see a clear improvement for all approaches. As expected, the largest change is observed for the constant mean model, since it has learned to rely almost entirely on observed data. We also observe a clear overestimation of the probabilities for the 20\,m/s and 30\,m/s thresholds. For the NWP mean model, we see a general reduction in the overestimation. For the neural mean approach, we see a general improvement in the calibration, especially for the 20\,m/s and 30\,m/s thresholds. The right-hand panel in Fig.~\ref{fig:mean_exp_pit} also reflects the observed changes. 

These results highlight the importance of providing an appropriate mean function to the model's prior. We have observed that, in our problem, including NWP information is crucial to obtain results comparable or better than the baseline, and that it is also important that the provided mean function is unbiased, thus requiring a sort of post-processing step applied to the raw NWP information. We note that while these results are true for the case of wind modeling, different outcomes could be obtained in other applications like temperature, where the observations are much more representative of the surroundings and an appropriate prior mean is therefore of less importance.

\begin{figure*}[t]
\includegraphics[width=\textwidth]{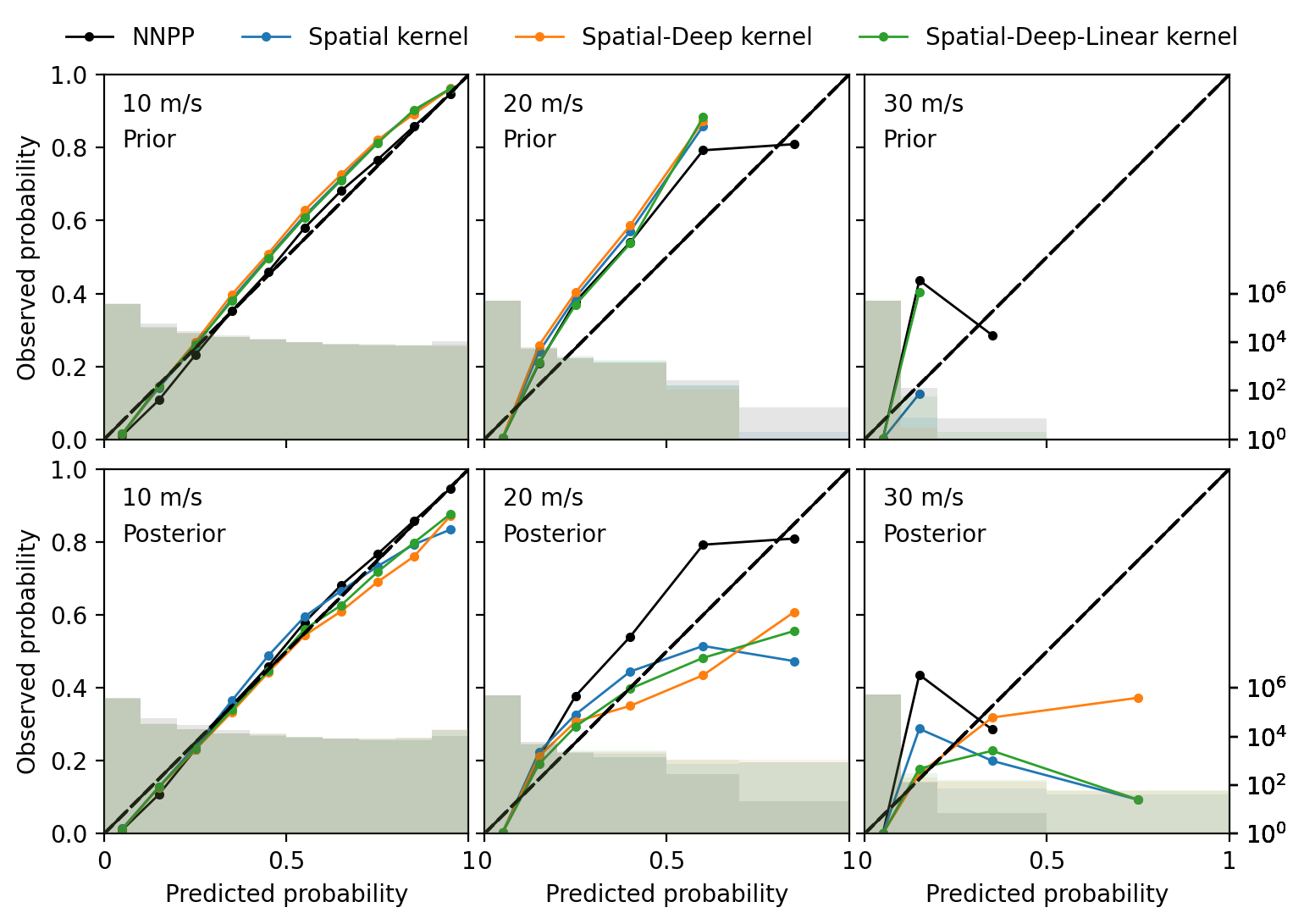}
\caption{Reliability diagrams for the covariance modeling experiments, with columns representing the different exceedance thresholds (10, 20 and 30\,$ms^{-1}$) and the first and second rows are for prior and posterior predictions respectively. The secondary axis shows the log-scaled number of occurrences for each bin.}
\label{fig:cov_exp_rel}
\end{figure*}

\begin{figure*}[t]
\includegraphics[width=\textwidth]{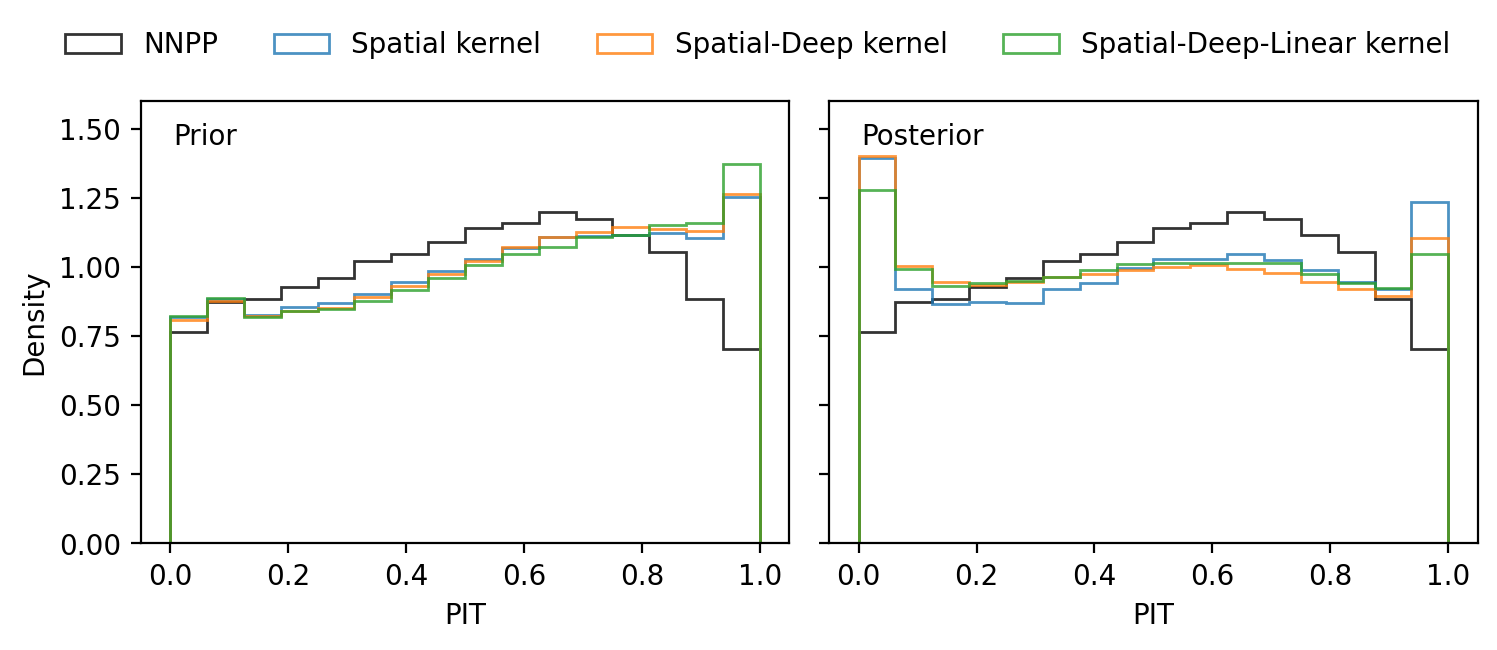}
\caption{cPIT histograms for the covariance modeling experiments, with the results for the prior and posterior predictions on the left and right hand side respectively. Perfect calibration would be indicated by a uniform histogram.}
\label{fig:cov_exp_pit}
\end{figure*}

\subsection{Covariance modeling}
\label{sec:covexp}

For the covariance function, we experimented with three variants of increasing complexity. We start with a simple spatial kernel $k_S: \mathbb{R}^3 \rightarrow \mathbb{R}$ as an RBF kernel taking as inputs easting and northing coordinates and elevation. We parameterize it as an automatic relevance determination kernel, which means that we have one length-scale for each input dimension. In practice, this means that some dimensions will be more important than others. We let the model learn both $\boldsymbol{l} \in \mathbb{R}^3$ and the variance parameter $\sigma^2$ (starting with some sensible defaults). 

The second variant consists of the product combination of the spatial kernel and a deep kernel, $k_{SD}(x, x') = k_{S}(x, x') \cdot k_{D}(x, x')$. Their combination was motivated by the shortcomings of the individual components. While the spatial kernel $k_{S}$ is likely too simplistic, a deep kernel can be susceptible to overfitting by wrongly learning spurious correlations over vast distances. Therefore, the covariance is forced to zero after a certain distance, preventing the model from picking up spurious correlations while still learning useful nonlinear relationships. We note that one could take advantage of this by using sparse computations as shown in \citet{furrer_covariance_2006}. However, this was not considered in this study.

The third variant $k_{SD}(x, x') = k_{S}(x, x') \cdot k_{D}(x, x') \cdot k_{L}(x, x')$ includes a linear scaling component on top of the spatial deep kernel. This allows us to learn a non-stationary component of the covariance structure, specifically a scaling term that depends on $x$ and $x'$ and not solely on their distance. 

For all variants, the same mean function is used, namely the neural mean function. 

The models considered for this experiment are shown in the third, fourth and fifth row of Fig.~\ref{fig:scorecard}. For prior predictions, we observe no significant difference in performance, with the exception of a small improvement for the $k_{SDL}$ kernel, which is on par with the baseline if no approximation is used. The difference between $k_{SDL}$ and the other two kernels that perform equally is due to the fact that $k_{SDL}$ is nonstationary, and is therefore able to learn an input-dependent prior marginal variance. On the other hand, stationary kernels have a fixed marginal variance for the prior predictive distribution. 

Figure~\ref{fig:cov_exp_rel} shows realiability diagrams for the prior and posterior prediction in the top and bottom row respectively. In general, results appear quite similar between the different models. We observe a general tendency to underestimate probabilities for prior predictions, whereas the opposite happens for posteriors although they are closer to being well calibrated. We should note that the number of occurrences for the second and third thresholds is very small for large probabilities, which makes the observed miscalibration less significant. Figure~\ref{fig:cov_exp_pit} confirms the same tendency: prior predictions are negatively biased and posteriors are not.

We observe that the largest improvement comes from the combination of the spatial kernel with the deep kernel, whereas the addition of the linear scaling does not change our results in a significant way.

\subsection{Approximation error}

Figure~\ref{fig:scorecard} also shows the effect of the RFF approximation on the quality of the predictions. In general, we observe that only posterior predictions suffer significantly from the approximation, whereas almost no effect is observed for prior predictions. We believe this might be explained by the fact that the prior function is smoother than the posterior and is therefore easier to approximate. As expected, increasing the number of Fourier features reduces the approximation error, with the trade-off of more expensive computation.

\section{Case study}
A qualitative analysis of the forecasts for a significant event is now provided. The predictions are generated by a model that integrates the Neural mean and the Spatial-Deep-Linear kernel, leveraging an RFF approximated kernel with 2048 Fourier features.

\begin{figure*}[t]
\includegraphics[width=\textwidth]{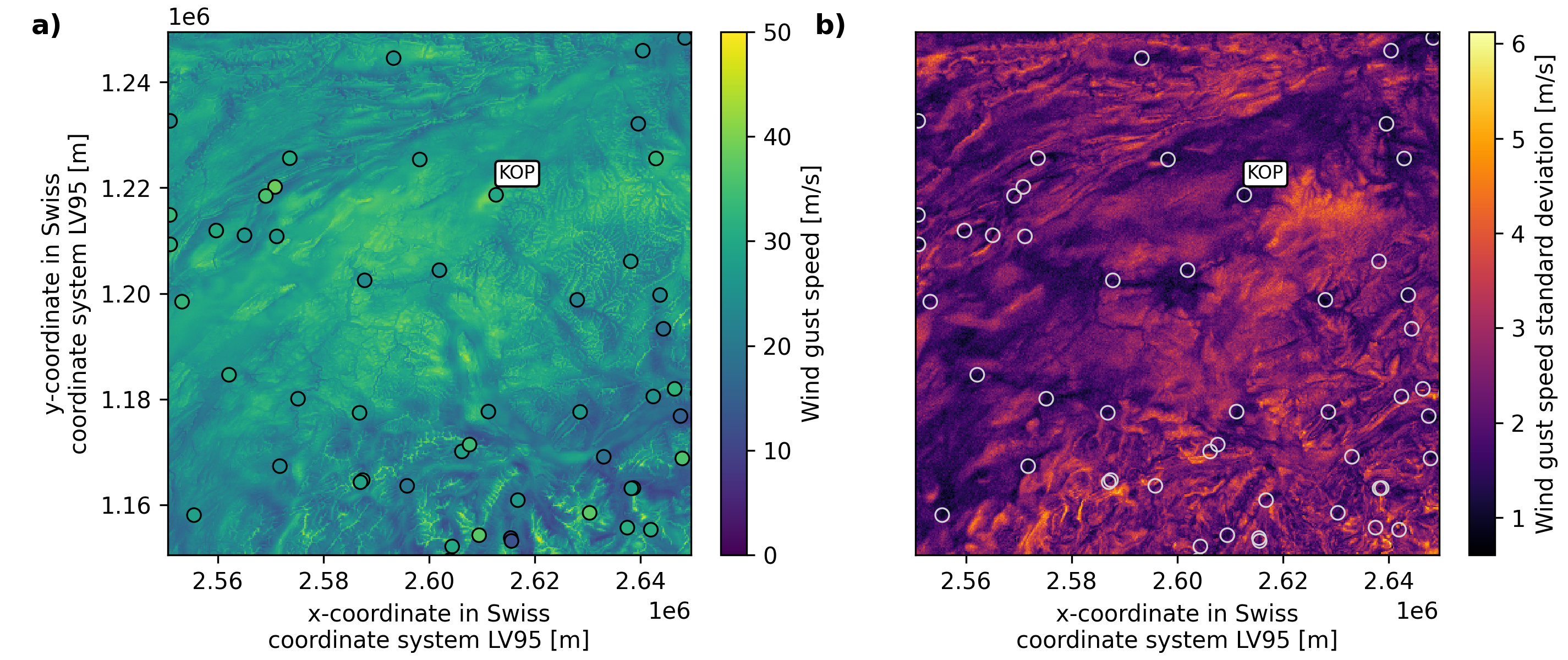}
\caption{\textbf{a.} The ensemble mean of the Mathis storm event's maximum wind speed, along with the measured station values shown in circles. \textbf{b}. The ensemble standard deviation of the Mathis storm event's maximum wind speed, with the location of the measuring stations shown with circles.}
\label{fig:mathis_maximum}
\end{figure*}

We examine the event that took place between March 31 and April 1, 2023, coinciding with the passage of storm Mathis across Switzerland. Our focus is on the northeastern part of Switzerland, which bore the brunt of the storm's impact, and we show predictions on a grid with a horizontal resolution of 250 meters. Figure~\ref{fig:mathis_maximum} illustrates the average predictions and standard deviation of the maximum wind gust predicted by each member of the ensemble during the event, alongside the observed values represented by circles. Overall, there appears to be strong agreement between the forecasts and the actual observations, displaying a realistic spatial pattern. The spatial pattern of the standard deviation is influenced by the distance from the measurement stations, particularly in flat regions. The explanation becomes more complex in areas with varied topography, where factors such as elevation variations and other geomorphological characteristics play an important role.

\begin{figure*}[t]
\includegraphics[width=\textwidth]{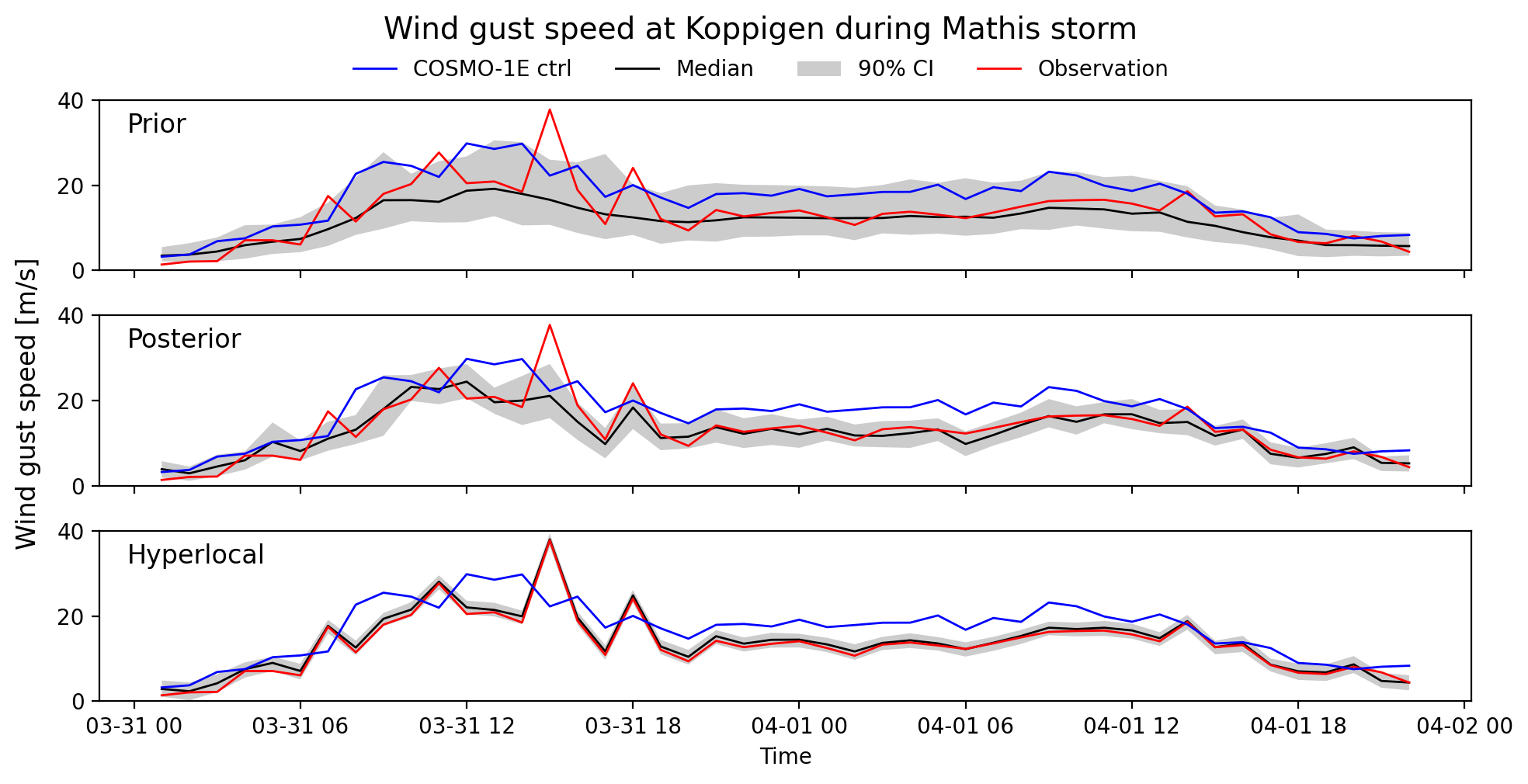}
\caption{Meteogram of the predicted and observed wind gust at the Koppigen station (KOP) during the Mathis storm event, using the approximated RFF covariance. The prior prediction (top) doesn't seem to capture hourly variability and has a relatively large spread throughout the event. On the other hand, the posterior prediction (middle) captures more  of the variability and does so with a sharper forecast. As one would expect, the hyperlocal prediction (bottom) closely follows the observed value, with an uncertainty corresponding to the observational noise.}
\label{fig:koppigen}
\end{figure*}

Figure~\ref{fig:koppigen} shows the meteograms for the Koppigen measuring station, where the maximum wind gust reached 136\,km h$^{-1}$ during the event, marking the highest value recorded in the area. Similarly to the diagram in Fig.~\ref{fig:scorecard}, we present results for three forecast categories: prior, posterior, and hyperlocal predictive distribution. In contrast to the prior predictions, the posterior predictive distribution seems narrower and somewhat more precise, indicating the beneficial impact of the information sourced from neighboring stations. Nevertheless, it is evident that certain high-intensity peaks are overlooked, possibly due to their highly localized nature and related unpredictability. Naturally, when examining the hyperlocal predictive distribution, these peaks are accurately represented.

\begin{figure*}[t]
\includegraphics[width=\textwidth]{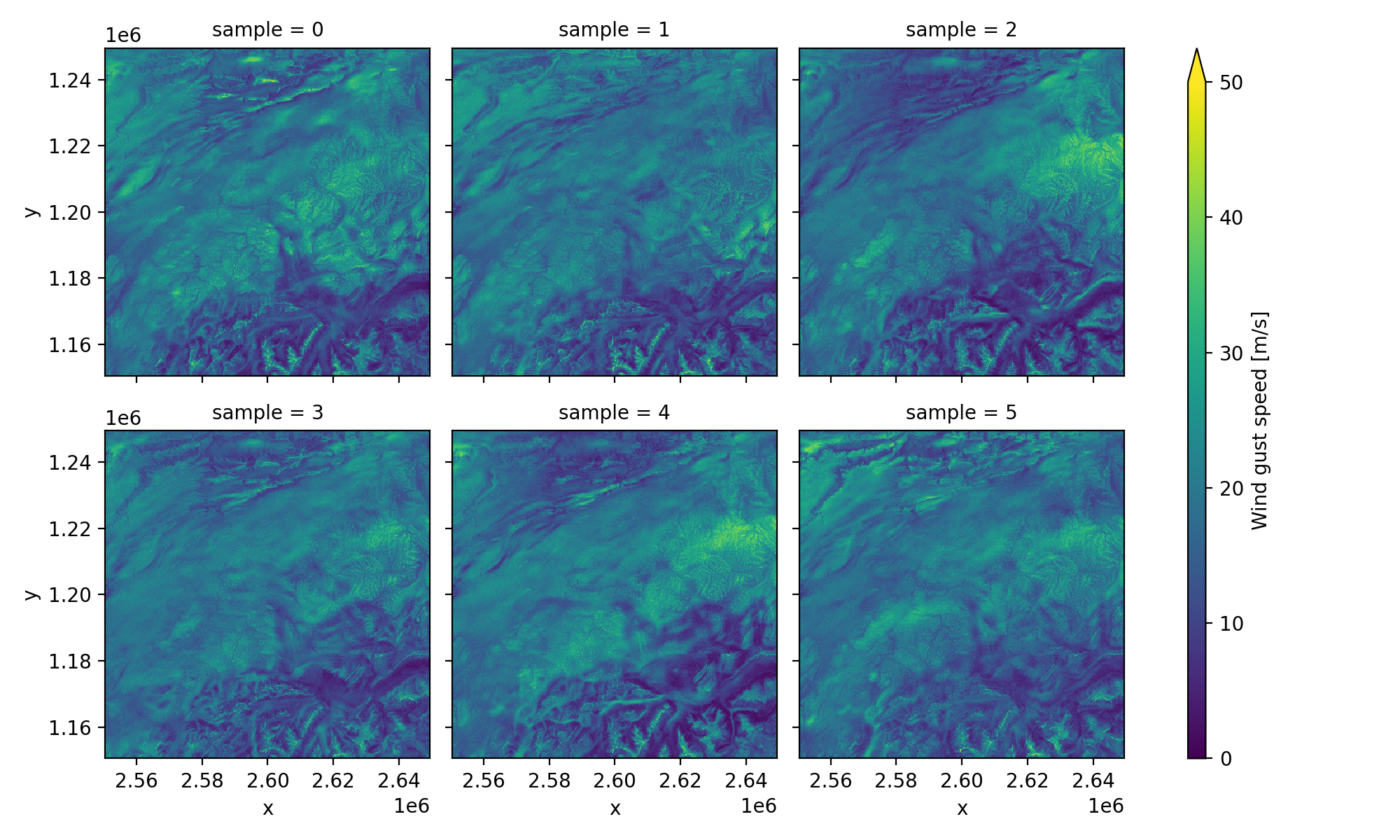}
\caption{A set of generated realizations from the posterior predictive distribution for the north-west region of Switzerland, during the Mathis storm event. Each realization displays a realistic spatial structure while still providing a good degree of variability on several spatial scales.}
\label{fig:realizations}
\end{figure*}

\begin{figure}[t]
\includegraphics[width=19pc]{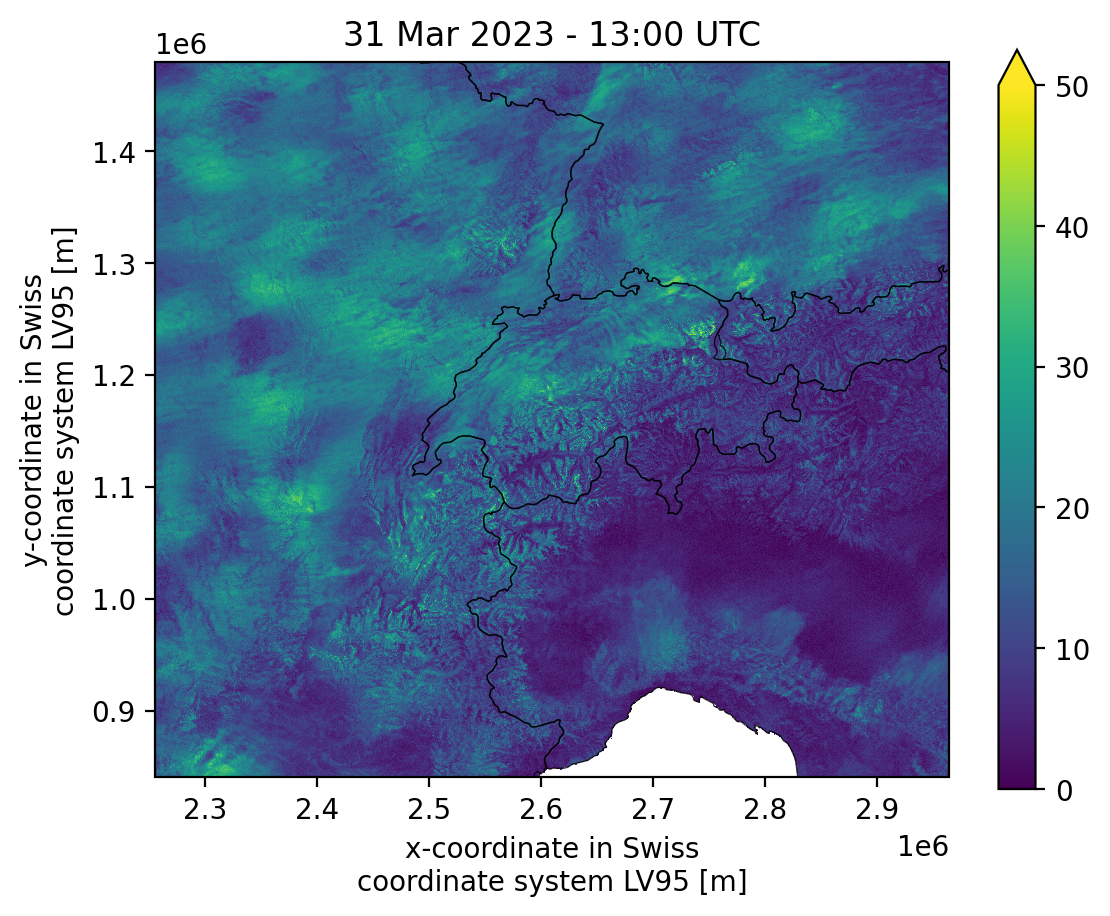}
\caption{A single generated realization from the posterior predictive distribution for the Swiss radar domain, during the Mathis storm event. The generated field displays a realistic spatial structure. The field covers a domain spanning 710\,km in longitude and 640\,km in latitude, and with a resolution of 250 meters it totals approximately 7.2 million pixels. Note that the real size of the grid cells is smaller than pixels in the image, resulting in some higher values that are barely visible to the eye.}
\label{fig:mathis_inca}
\end{figure}

The RFF approximation, in combination with the pathwise conditioning approach, allows us to efficiently sample from the posterior predictive distribution. Figure~\ref{fig:realizations} shows a series of random realizations at a single timestep during the event. All realizations have a realistic spatial structure coherent with the used topographical features. However, it is important to consider that this covariance structure was learned in a data-driven way and from sparse observations, and the learned covariance structure was approximated. While these realizations certainly look plausible, there is no guarantee that they are physically coherent. Finally, Figure~\ref{fig:mathis_inca} demonstrates the scalability of our sampling approach by generating a coherent realization on a 250\,m resolution grid with 7.2\,million points. Sampling 51 of these realizations with the covariance approximated by 2048 Fourier features takes roughly 7 seconds on our hardware (a single NVIDIA A-100 GPU). It should be noted that when compared with the raw NWP predictions from \ref{fig:mathis_inca_nwp}, some of the large scale anisotropic structure of the wind flow is not well captured. We speculate that this result relates to predictability: if on one hand topographically forced wind is more predictable, making it easier to learn covariance structure at small scales, on the other the dependencies related to these large scale features are less systematic and subject to noise, and therefore harder to learn.

\section{Conclusions and outlook}
This study presented a methodology for accurately representing sub-kilometer surface wind gust through a combination of ML methods, namely GPs and NNs. We have shown the added value of including surface measurements compared to simply applying a post-processing model to NWP forecasts. We addressed the computational challenges of GPs by leveraging state-of-the-art techniques for efficient sampling, which allows our models to generate spatially coherent realizations of wind fields that are also marginally calibrated.
Our experiments, while demonstrating the flexibility our the proposed methodology, highlight the importance of providing good priors to GPs. We speculate that this could be crucial for achieving incremental improvements.

\subsection{Related work}
The proposed methodology shares similarities with the approaches of neural processes \citep[NP]{garnelo_neural_2018}, in that both combine ideas from GPs with NNs to model stochastic processes, and the metalearning algorithm used here is analogous to what is commonly used in NPs. However, there are notable differences. While NPs directly estimate the entire probabilistic mapping through a NN, our method maintains an explicit GP structure with a learnable mean and covariance function, and the Bayesian rule is explicitly computed instead of being learned in a latent space. 
Generally speaking, NPs offer more flexibility compared to GPs: they can better handle spatio-temporal data in their convolutional versions, and they are able to learn more complex, non-stationary covariance structures. They also have better scaling capabilities, as computations linearly scale with the number of input points.
For example, \citet{vaughan_convolutional_2022} proposed the use of ConvCNPs for a local climate downscaling task, showing better performance compared to standard GPs, especially for meteorological fields with complex covariance structures such as precipitation. Similar results where obtained by \citet{scholz_sim2real_2023}. The main limitation of this approach is that the spatio-temporal structure is not considered as covariance is not computed. This issue was addressed in another work by \citet{markou_practical_2022}, which introduced the ConvGNP architecture, capable of directly estimating the predictive covariance of a multivariate Gaussian and therefore allowing one to draw coherent samples. \citet{andersson_environmental_2023} is another relevant contribution in this area, based on the ConvGNP.
While bringing the aforementioned flexibility and computational efficiency during inference, NPs also have some disadvantages compared to GPs. They need larger datasets and take longer to train, making GPs a better alternative in data-scarce situations, and the richer covariance modeling comes with the disadvantage of being less explicit, not allowing for designing domain-specific components. In summary, what differentiates this work from similar NPs applications is that the GP structure is maintained while flexibility is still improved through the integration of NNs. Furthermore, we extended our modeling with RFF and pathwise conditioning, allowing more scalability for sampling coherent realizations on dense grids, which could in principle be a useful extension for ConvGNPs as well.

\subsection{Outlook}

We envisage that the following aspects could be further explored in future research:

\begin{itemize}
    \item Modeling the temporal covariance structure explicitly, by including a temporal covariance kernel function. This could bring potentially significant benefits when considering a finer temporal granularity, for instance 10\,minutes instead of hourly granularity as presented here;
    \item Explore the potential of the proposed methodology for nowcasting applications by developing an appropriate blending function controlled by the observational noise;
    \item Further improving the data transformation technique. For example, one might consider learning input-dependent transformations as proposed in \citet{maronas_transforming_2021};
    \item Extending current models to handle vector data, such as wind vectors.

\end{itemize}

Furthermore, an interesting question would be the direct comparison of the proposed methodology with NPs for spatio-temporal modeling tasks, allowing a more detailed discussion of the relative advantages and disadvantages.

%

%

\clearpage

\acknowledgments
We thank Thomas Pinder and Daniel Dodd for our fruitful discussions and for their work on the GPJax library, which significantly helped the project. We thank many of the MeteoSwiss colleagues for their helpful comments and feedback. Author Zanetta is supported by MeteoSwiss and ETH Zurich, authors Nerini and Buzzi are supported by MeteoSwiss, and author Moss is supported by the University of Cambridge. We also thank the Swiss National Supercomputing Centre (CSCS) in Lugano for its computing infrastructure.
%

\datastatement
The project’s GitHub repository is accessible online \href{https://github.com/frazane/swagp}{https://github.com/frazane/swagp}. The raw data used to train the models is free for research and education purposes, and can be accessed via the IDAweb portal at \url{https://www.meteoswiss.admin.ch/services-and-publications/service/weather-and-climate-products/data-portal-for-teaching-and-research.html}


\appendix
\appendixtitle{Implementation details}

\subsection{Data partitioning} \label{appendix:partitioning}
\begin{figure*}[t]
\includegraphics[width=\textwidth]{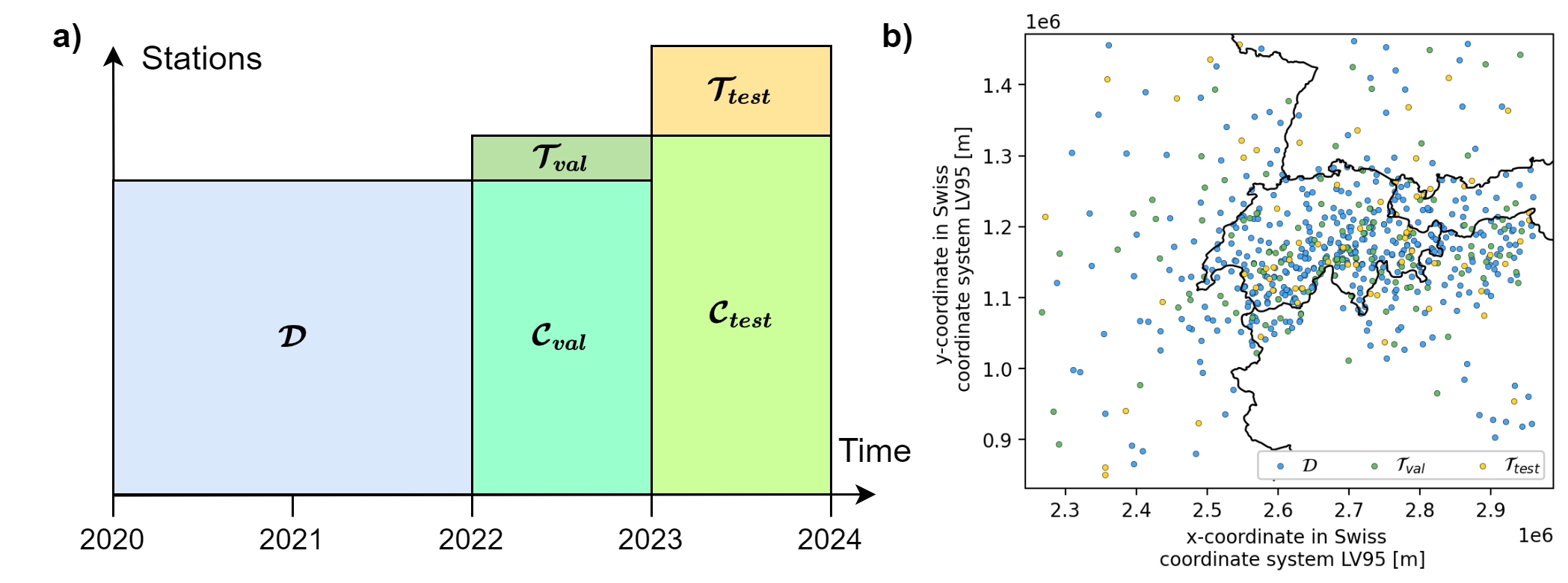}
\caption{\textbf{a.} Schematic of the data partitioning strategy. \textbf{b.} Spatial distribution of the $\boldsymbol{\mathcal{D}}$, $\boldsymbol{\mathcal{T}_{val}}$ and $\boldsymbol{\mathcal{T}_{test}}$ sets.}
\label{fig:partitioning}
\end{figure*}

For partitioning the stations, we used stratified random sampling, by which the stations were first grouped by their 99$^{th}$ quantile values. This was an easy-to-implement approach to ensure that heterogeneous conditions in the surface stations (hyperlocal conditions, type of measuring device, height of measuring device, etc.) were equally represented in the different sets, with a focus on the tails of the distributions. The time partitioning is sequential, with years 2020 and 2021 for the training set, 2022 for validation, and 2023 for the test set.
As explained in the main text, the evaluation of the validation and test set is performed on a set $\mathcal{T}$ of station measurements, with other sets of station measurements $\mathcal{C}$ used as context. In the case of validation, stations from the training set $\boldsymbol{\mathcal{D}}$ are used as context. In the case of testing, stations from both training $\boldsymbol{\mathcal{D}}$ and validation targets $\boldsymbol{\mathcal{T}_{val}}$ are used as context. A visual representation of the partitioning strategy is shown in Figure~\ref{fig:partitioning}a. Figure~\ref{fig:partitioning}b shows the spatial distribution of $\boldsymbol{\mathcal{D}}$, $\boldsymbol{\mathcal{T}_{val}}$ and $\boldsymbol{\mathcal{T}_{test}}$. While this partitioning strategy does not create strictly independent datasets, we believe it is the best possible solution. It reflects the real-world conditions in which the presented methodology would be applied, where the stations used during training will be used as context for the predicted fields. Moreover, the same approach was already used in related work by \citet{scholz_sim2real_2023}.

\subsection{Data transformation} \label{appendix:transformation}

One of the basic assumptions of GPs is that the random variable of interest is normally distributed. In real-world applications, this is often not the case, and the use of GPs with non-conjugate posteriors is an active area of research. While some theoretically principled approaches exist to deal with these situations, such as Markov Chain Monte Carlo methods, Laplace approximation, or Variational Inference, we opted for a simpler approach and tested it empirically on our problem. Our approach consists of transforming the original data to a Gaussian distribution using a parametric transformation that is also bijective and differentiable.

We have defined a parametric transformation for our data of the form
$$
z = - \frac{\log{\frac{a}{y} - c}}{b}
$$
with its corresponding inverse transformation
$$
y = \frac{a}{c + e^{-bz}}
$$
where parameters $a=4.66$ and $b=0.74$ and $c=0.08$ were found via curve fitting on a corresponding empirical quantile transformation. A Jupyter notebook illustrating the procedure is included in the code repository. Compared to using the quantile transformation directly, we have observed that this parametric version behaved more robustly, especially towards the right tail of the distribution where there is a very low density of points. Another advantage of this parametric transformation is that it allows us to easily compute its derivative, $\frac{d z}{d y} = \frac{a}{by(a - cy)}$, which can be used to scale the observational noise to compensate for the deformation of the density distribution. On the other hand, using a homoscedastic observational noise in the transformed space would result in a heteroscedastic observational noise in the original space, especially for large values of $y$, after applying the inverse transformation to sampled values.

\subsection{Evaluation metrics and tools} \label{appendix:evaluation}

We evaluate the performance of our models using standard tools for probabilistic forecast verification. Specifically, we have used the CRPS, as defined by
$$
\text{CRPS}(F, y) = \int_{\mathbb{R}}[F(x)-\mathbb{1}\{y \le x\}]^2 dx,
$$
to evaluate sharpness and calibration of the forecasts. To put less emphasis on small values of wind gusts, which are practically irrelevant for our application, we have decided to opt for a threshold-weighted version of the CRPS \citep[twCRPS]{allen_weighted_2022}, and considered the twCRPS for a threshold value of 4\,$ms^{-1}$.
We have also used two other tools to evaluate calibration. The reliability diagram, which shows how well the predicted probabilities of an event correspond to their observed frequencies, can be used to assess calibration with respect to specific thresholds. The PIT histogram \citep{gneiting_probabilistic_2007}, can be used to assess the overall calibration and determines whether the random variable of interest is indeed sampled from the predicted distribution, in which case the PIT would be uniformly distributed. As with the twCRPS, we opted for a version of the PIT histogram, the cPIT \citep{allen_weighted_2022}, that emphasizes calibration above the same threshold.

\subsection{Modeling details} \label{appendix:training}

\subsubsection{Architectures, optimization and predictors}

The neural network used for the mean function was a fully connected network with 2 layers and 32 units each, using the hyperbolic tangent activation function. The same architecture was used for the deep kernel $k_D$, with the only difference that it outputs a two-dimensional vector (which is passed to the base kernel) instead of a one-dimensional vector. All models were trained with a batch size of 64 timesteps/tasks and used the Adabelief optimizer \citep{zhuang_adabelief_2020} with a learning rate of $10^{-4}$. During training, the repeated presence of the same set of stations at all steps could potentially lead to co-adaptation/overfitting. To mitigate this risk, each sampled task contained a different random set of 128 stations. Additionally, the CRPS for marginal predictions was computed as a validation loss and used to diagnose overfitting to the training dataset and guided us in the choice of the models' hyperparameters.
Additional details about the models can be inspected in the manuscript's accompanying GitHub repository.

\begin{table}
    \centering
    \begin{tabular}{rr}
        Name & Type \\
        \midrule
         Surface wind speed of gust & Dynamic \\
         Surface wind speed of gust ($t - 1$ hour) & Dynamic \\
         Surface wind speed of gust ($t + 1$ hour) & Dynamic \\
         Change in wind speed of gust ($t - 1$ to $t + 1$) & Dynamic \\
         Sx \citep{winstral_spatial_2002} & Dynamic/Static \\
         Sine component of the hour of the day & Temporal \\
         Cosine component of the hour of the day & Temporal \\
         TPI (500\,m scale) & Static \\
         TPI (2000\,m scale) & Static \\
         Model elevation difference & Static \\
         West-East derivative (2000\, scale) & Static \\
         South-North derivative (2000\, scale) & Static \\
         \bottomrule
    \end{tabular}
    \caption{Predictors used by the NNPP baseline model.}
    \label{tab:nnpp_predictors}
\end{table}

\begin{table}
    \centering
    \begin{tabular}{rr}
        Name & Type \\
        \midrule
         Surface wind speed of gust & Dynamic \\
         Surface wind speed of gust ($t - 1$ hour) & Dynamic \\
         Surface wind speed of gust ($t + 1$ hour) & Dynamic \\
         Sx \citep{winstral_spatial_2002} & Dynamic/Static \\
         Sine component of the hour of the day & Temporal \\
         Cosine component of the hour of the day & Temporal \\
         TPI (500\,m scale) & Static \\
         TPI (2000\,m scale) & Static \\
         Model elevation difference & Static \\
         West-East derivative (2000\, scale) & Static \\
         South-North derivative (2000\, scale) & Static \\
        \bottomrule
    \end{tabular}
    \caption{Predictors used by the neural mean.}
    \label{tab:neural_mean_predictors}
\end{table}

\begin{table}
    \centering
    \begin{tabular}{rr}
        Name & Type \\
        \midrule
         Elevation & Static \\
         Easting in Swiss coordinates & Static \\
         Northing in Swiss coordinates & Static \\
         \bottomrule
    \end{tabular}
    \caption{Predictors used by the spatial kernel.}
    \label{tab:spatial_kernel_predictors}
\end{table}

\begin{table}
    \centering
    \begin{tabular}{rr}
        Name & Type \\
        \midrule
         Eastward wind component & Dynamic \\
         Northward wind component & Dynamic \\
         Model elevation difference & Static \\
         Elevation & Static \\
         Easting in Swiss coordinates & Static \\
         Northing in Swiss coordinates & Static \\
         West-East derivative (2000\, scale) & Static \\
         South-North derivative (2000\, scale) & Static \\
         \bottomrule
    \end{tabular}
    \caption{Predictors used by the deep kernel.}
    \label{tab:deep_kernel_predictors}
\end{table}

\begin{table}
    \centering
    \begin{tabular}{rr}
        Name & Type \\
        \midrule
         Surface wind speed of gust & Dynamic \\
         Change in wind speed of gust ($t - 1$ to $t + 1$) & Dynamic \\
         Sine component of the hour of the day & Temporal \\
         Cosine component of the hour of the day & Temporal \\
         Model elevation difference & Static \\
         TPI (2000\,m scale) & Static \\
         West-East derivative (2000\, scale) & Static \\
         South-North derivative (2000\, scale) & Static \\
         \bottomrule
    \end{tabular}
    \caption{Predictors used by the linear scaling component of the Spatial-Deep-Linear kernel.}
    \label{tab:linear_kernel_predictors}
\end{table}

\subsubsection{Parameters constraints}
When training GPs models, one must take into consideration that some parameters can only take positive values, such as a kernel's variance and length-scale parameters or the observational noise. To ensure that these parameters do not become negative during optimization, we have considered two common approaches. The first approach was the same implemented in \citet{pinder_gpjax_2022}, which uses bijective transformations to map parameters to and from an unconstrained space where the gradient updates are applied. This approach is very general and can accommodate a large variety of constraints over the model parameters. Since we were only interested in ensuring that parameters remain strictly positive, we opted for another approach that simply consists of modifying the optimization updates to keep parameters above zero. The constraint function $f$ is then defined as
\begin{equation}
    f(p, u) = 
    \begin{cases} 
        -p + \delta & \text{if } p + u < \delta \\
        u & \text{otherwise}
    \end{cases}
\end{equation}
where $p$ is the current value of a parameter, $u$ is the computed update (already scaled by the learning rate) based on the gradient of the loss function, and $\delta$ is a small positive value.

\subsection{Libraries and computational resources} \label{appendix:libraries}

Our approach centers around JAX \citep{bradbury_jax_2018}, a  framework designed for high-performance computing and ML, which includes automatic differentiation features. We incorporated modified code from the GPJax \citep{pinder_gpjax_2022} library. We used a single NVIDIA Tesla A-100 GPU for model training, significantly reducing the training time, thanks to JAX's efficient just-in-time compilation. For instance, training the model with the neural mean and the spatial deep kernel took us around one minute. We conducted our computations at the Swiss Centre for Scientific Computing (CSCS).

\subsection{Additional figures}
\begin{figure}[h]
    \includegraphics[width=19pc]{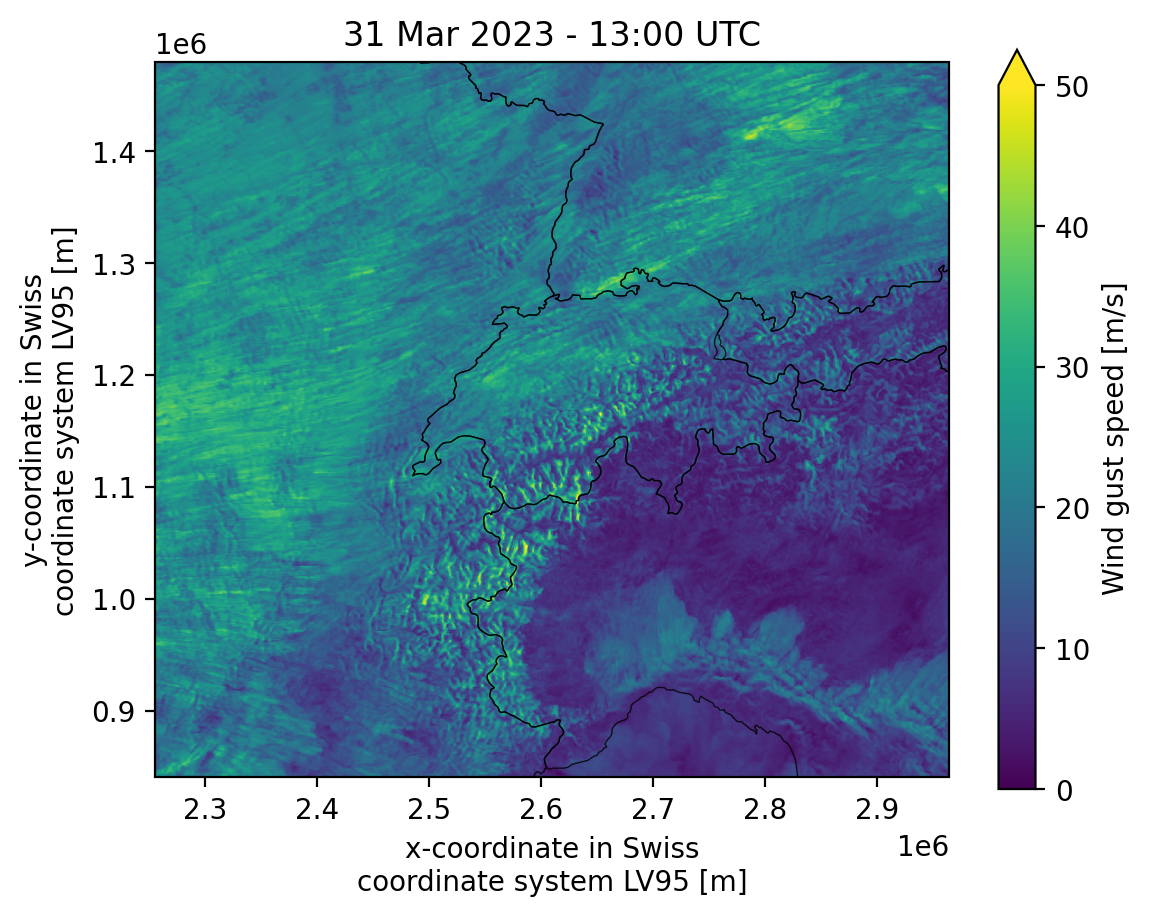}
    \caption{A COSMO-1E control prediction during the Mathis event.}
    \label{fig:mathis_inca_nwp}
\end{figure}

\newpage




%



\bibliographystyle{ametsocV6}
\bibliography{references}

\begin{thebibliography}{35}
\providecommand{\natexlab}[1]{#1}
\providecommand{\url}[1]{\texttt{#1}}
\renewcommand{\UrlFont}{\rmfamily}
\providecommand{\urlprefix}{URL }
\expandafter\ifx\csname urlstyle\endcsname\relax
  \providecommand{\doi}[1]{https://doi.org/\discretionary{}{}{}#1}\else
  \providecommand{\doi}{https://doi.org/\discretionary{}{}{}\begingroup \urlstyle{rm}\Url}\fi
\providecommand{\eprint}[2][]{\url{#2}}

\bibitem[{Allen et~al.(2022)Allen, Bhend, Martius,, and Ziegel}]{allen_weighted_2022}
Allen, S., J.~Bhend, O.~Martius, and J.~Ziegel, 2022: Weighted verification tools to evaluate univariate and multivariate forecasts for high-impact weather events. arXiv, \urlprefix\url{http://arxiv.org/abs/2209.04872}, \doi{10.48550/arXiv.2209.04872}.

\bibitem[{Andersson et~al.(2023)}]{andersson_environmental_2023}
Andersson, T.~R., and Coauthors, 2023: Environmental sensor placement with convolutional {Gaussian} neural processes. \textit{Environmental Data Science}, \textbf{2}, e32, \doi{10.1017/eds.2023.22}.

\bibitem[{Bernhardt et~al.(2009)Bernhardt, Zängl, Liston, Strasser,, and Mauser}]{bernhardt_using_2009}
Bernhardt, M., G.~Zängl, G.~E. Liston, U.~Strasser, and W.~Mauser, 2009: Using wind fields from a high‐resolution atmospheric model for simulating snow dynamics in mountainous terrain. \textit{Hydrological Processes}, \textbf{23~(7)}, 1064--1075, \doi{10.1002/hyp.7208}.

\bibitem[{Bradbury et~al.(2018)}]{bradbury_jax_2018}
Bradbury, J., and Coauthors, 2018: {JAX}: composable transformations of {Python}+{NumPy} programs. \urlprefix\url{http://github.com/google/jax}.

\bibitem[{Christianson et~al.(2023)Christianson, Pollyea,, and Gramacy}]{christianson_traditional_2023}
Christianson, R.~B., R.~M. Pollyea, and R.~B. Gramacy, 2023: Traditional kriging versus modern {Gaussian} processes for large‐scale mining data. \textit{Statistical Analysis and Data Mining: The ASA Data Science Journal}, \textbf{16~(5)}, 488--506, \doi{10.1002/sam.11635}.

\bibitem[{Clark et~al.(2004)Clark, Gangopadhyay, Hay, Rajagopalan,, and Wilby}]{clark_schaake_2004}
Clark, M., S.~Gangopadhyay, L.~Hay, B.~Rajagopalan, and R.~Wilby, 2004: The {Schaake} {Shuffle}: {A} {Method} for {Reconstructing} {Space}–{Time} {Variability} in {Forecasted} {Precipitation} and {Temperature} {Fields}. \textit{Journal of Hydrometeorology}, \textbf{5~(1)}, 243--262, \doi{10.1175/1525-7541(2004)005<0243:TSSAMF>2.0.CO;2}.

\bibitem[{Demaeyer et~al.(2023)}]{demaeyer_euppbench_2023}
Demaeyer, J., and Coauthors, 2023: The {EUPPBench} postprocessing benchmark dataset v1.0. \textit{Earth System Science Data}, \textbf{15~(6)}, 2635--2653, \doi{10.5194/essd-15-2635-2023}.

\bibitem[{Dujardin and Lehning(2022)Dujardin, and Lehning}]{dujardin_windtopo_2022}
Dujardin, J., and M.~Lehning, 2022: Wind‐{Topo}: {Downscaling} near‐surface wind fields to high‐resolution topography in highly complex terrain with deep learning. \textit{Quarterly Journal of the Royal Meteorological Society}, \textbf{148~(744)}, 1368--1388, \doi{10.1002/qj.4265}.

\bibitem[{Furrer et~al.(2006)Furrer, Genton,, and Nychka}]{furrer_covariance_2006}
Furrer, R., M.~G. Genton, and D.~Nychka, 2006: Covariance {Tapering} for {Interpolation} of {Large} {Spatial} {Datasets}. \textit{Journal of Computational and Graphical Statistics}, \textbf{15~(3)}, 502--523, \doi{10.1198/106186006X132178}.

\bibitem[{Gardner et~al.(2021)Gardner, Pleiss, Bindel, Weinberger,, and Wilson}]{gardner_gpytorch_2021}
Gardner, J.~R., G.~Pleiss, D.~Bindel, K.~Q. Weinberger, and A.~G. Wilson, 2021: {GPyTorch}: {Blackbox} {Matrix}-{Matrix} {Gaussian} {Process} {Inference} with {GPU} {Acceleration}. arXiv, \urlprefix\url{http://arxiv.org/abs/1809.11165}, arXiv:1809.11165 [cs, stat], \doi{10.48550/arXiv.1809.11165}.

\bibitem[{Garnelo et~al.(2018)Garnelo, Schwarz, Rosenbaum, Viola, Rezende, Eslami,, and Teh}]{garnelo_neural_2018}
Garnelo, M., J.~Schwarz, D.~Rosenbaum, F.~Viola, D.~J. Rezende, S.~Eslami, and Y.~W. Teh, 2018: Neural processes. \textit{arXiv preprint arXiv:1807.01622}.

\bibitem[{Gneiting et~al.(2007)Gneiting, Balabdaoui,, and Raftery}]{gneiting_probabilistic_2007}
Gneiting, T., F.~Balabdaoui, and A.~E. Raftery, 2007: Probabilistic {Forecasts}, {Calibration} and {Sharpness}. \textit{Journal of the Royal Statistical Society Series B: Statistical Methodology}, \textbf{69~(2)}, 243--268, \doi{10.1111/j.1467-9868.2007.00587.x}.

\bibitem[{Hemri et~al.(2014)Hemri, Scheuerer, Pappenberger, Bogner,, and Haiden}]{hemri_trends_2014}
Hemri, S., M.~Scheuerer, F.~Pappenberger, K.~Bogner, and T.~Haiden, 2014: Trends in the predictive performance of raw ensemble weather forecasts. \textit{Geophysical Research Letters}, \textbf{41~(24)}, 9197--9205, \doi{10.1002/2014GL062472}.

\bibitem[{Liu et~al.(2020)Liu, Ong, Shen,, and Cai}]{liu_when_2020}
Liu, H., Y.-S. Ong, X.~Shen, and J.~Cai, 2020: When {Gaussian} {Process} {Meets} {Big} {Data}: {A} {Review} of {Scalable} {GPs}. \textit{IEEE Transactions on Neural Networks and Learning Systems}, \textbf{31~(11)}, 4405--4423, \doi{10.1109/TNNLS.2019.2957109}.

\bibitem[{Lundquist et~al.(2019)Lundquist, Hughes, Gutmann,, and Kapnick}]{lundquist_our_2019}
Lundquist, J., M.~Hughes, E.~Gutmann, and S.~Kapnick, 2019: Our {Skill} in {Modeling} {Mountain} {Rain} and {Snow} is {Bypassing} the {Skill} of {Our} {Observational} {Networks}. \textit{Bulletin of the American Meteorological Society}, \textbf{100~(12)}, 2473--2490, \doi{10.1175/BAMS-D-19-0001.1}.

\bibitem[{Markou et~al.(2022)Markou, Requeima, Bruinsma, Vaughan,, and Turner}]{markou_practical_2022}
Markou, S., J.~Requeima, W.~P. Bruinsma, A.~Vaughan, and R.~E. Turner, 2022: Practical {Conditional} {Neural} {Processes} {Via} {Tractable} {Dependent} {Predictions}. arXiv, \urlprefix\url{http://arxiv.org/abs/2203.08775}, \doi{10.48550/arXiv.2203.08775}.

\bibitem[{Maroñas et~al.(2021)Maroñas, Hamelijnck, Knoblauch,, and Damoulas}]{maronas_transforming_2021}
Maroñas, J., O.~Hamelijnck, J.~Knoblauch, and T.~Damoulas, 2021: Transforming {Gaussian} {Processes} {With} {Normalizing} {Flows}. arXiv, \urlprefix\url{http://arxiv.org/abs/2011.01596}, arXiv:2011.01596 [cs], \doi{10.48550/arXiv.2011.01596}.

\bibitem[{Matthews et~al.(2017)Matthews, Wilk, Nickson, Fujii, Boukouvalas, Le\{{\textbackslash}'o\}n-Villagr\{{\textbackslash}'a\}, Ghahramani,, and Hensman}]{matthews_gpflow_2017}
Matthews, A. G. d.~G., M.~v.~d. Wilk, T.~Nickson, K.~Fujii, A.~Boukouvalas, P.~Le\{{\textbackslash}'o\}n-Villagr\{{\textbackslash}'a\}, Z.~Ghahramani, and J.~Hensman, 2017: {GPflow}: {A} {Gaussian} {Process} {Library} using {TensorFlow}. \textit{Journal of Machine Learning Research}, \textbf{18~(40)}, 1--6.

\bibitem[{Mott and Lehning(2010)Mott, and Lehning}]{mott_meteorological_2010}
Mott, R., and M.~Lehning, 2010: Meteorological {Modeling} of {Very} {High}-{Resolution} {Wind} {Fields} and {Snow} {Deposition} for {Mountains}. \textit{Journal of Hydrometeorology}, \textbf{11~(4)}, 934--949, \doi{10.1175/2010JHM1216.1}.

\bibitem[{Patacchiola et~al.(2020)Patacchiola, Turner, Crowley, O'Boyle,, and Storkey}]{patacchiola_bayesian_2020}
Patacchiola, M., J.~Turner, E.~J. Crowley, M.~O'Boyle, and A.~Storkey, 2020: Bayesian {Meta}-{Learning} for the {Few}-{Shot} {Setting} via {Deep} {Kernels}. arXiv, \urlprefix\url{http://arxiv.org/abs/1910.05199}, arXiv:1910.05199 [cs, stat], \doi{10.48550/arXiv.1910.05199}.

\bibitem[{Pinder and Dodd(2022)Pinder, and Dodd}]{pinder_gpjax_2022}
Pinder, T., and D.~Dodd, 2022: {GPJax}: {A} {Gaussian} {Process} {Framework} in {JAX}. \textit{Journal of Open Source Software}, \textbf{7~(75)}, 4455, \doi{10.21105/joss.04455}.

\bibitem[{Rahimi and Recht(2007)Rahimi, and Recht}]{rahimi_random_2007}
Rahimi, A., and B.~Recht, 2007: Random {Features} for {Large}-{Scale} {Kernel} {Machines}. \textit{Advances in {Neural} {Information} {Processing} {Systems}}, Curran Associates, Inc., Vol.~20, \urlprefix\url{https://papers.nips.cc/paper_files/paper/2007/hash/013a006f03dbc5392effeb8f18fda755-Abstract.html}.

\bibitem[{Rasmussen and Williams(2006)Rasmussen, and Williams}]{rasmussen_gaussian_2006}
Rasmussen, C.~E., and C.~K.~I. Williams, 2006: \textit{Gaussian processes for machine learning}. Adaptive computation and machine learning, MIT Press, Cambridge, Mass, oCLC: ocm61285753.

\bibitem[{Rasp and Lerch(2018)Rasp, and Lerch}]{rasp_neural_2018}
Rasp, S., and S.~Lerch, 2018: Neural {Networks} for {Postprocessing} {Ensemble} {Weather} {Forecasts}. \textit{Monthly Weather Review}, \textbf{146~(11)}, 3885--3900, \doi{10.1175/MWR-D-18-0187.1}.

\bibitem[{Schefzik et~al.(2013)Schefzik, Thorarinsdottir,, and Gneiting}]{schefzik_uncertainty_2013}
Schefzik, R., T.~L. Thorarinsdottir, and T.~Gneiting, 2013: Uncertainty {Quantification} in {Complex} {Simulation} {Models} {Using} {Ensemble} {Copula} {Coupling}. \textit{Statistical Science}, \textbf{28~(4)}, \doi{10.1214/13-STS443}.

\bibitem[{Scholz et~al.(2023)Scholz, Andersson, Vaughan, Requeima,, and Turner}]{scholz_sim2real_2023}
Scholz, J., T.~R. Andersson, A.~Vaughan, J.~Requeima, and R.~E. Turner, 2023: {Sim2Real} for {Environmental} {Neural} {Processes}. arXiv, \urlprefix\url{http://arxiv.org/abs/2310.19932}, \doi{10.48550/arXiv.2310.19932}.

\bibitem[{Schraff et~al.(2016)Schraff, Reich, Rhodin, Schomburg, Stephan, Periáñez,, and Potthast}]{schraff_kilometrescale_2016}
Schraff, C., H.~Reich, A.~Rhodin, A.~Schomburg, K.~Stephan, A.~Periáñez, and R.~Potthast, 2016: Kilometre‐scale ensemble data assimilation for the {COSMO} model ({KENDA}). \textit{Quarterly Journal of the Royal Meteorological Society}, \textbf{142~(696)}, 1453--1472, \doi{10.1002/qj.2748}.

\bibitem[{Vannitsem(2018)}]{vannitsem_statistical_2018}
Vannitsem, S., Ed., 2018: \textit{Statistical postprocessing of ensemble forecasts}. Elsevier, Amsterdam, Netherlands.

\bibitem[{Vannitsem et~al.(2021)}]{vannitsem_statistical_2021}
Vannitsem, S., and Coauthors, 2021: Statistical {Postprocessing} for {Weather} {Forecasts}: {Review}, {Challenges}, and {Avenues} in a {Big} {Data} {World}. \textit{Bulletin of the American Meteorological Society}, \textbf{102~(3)}, E681--E699, \doi{10.1175/BAMS-D-19-0308.1}.

\bibitem[{Vaughan et~al.(2022)Vaughan, Tebbutt, Hosking,, and Turner}]{vaughan_convolutional_2022}
Vaughan, A., W.~Tebbutt, J.~S. Hosking, and R.~E. Turner, 2022: Convolutional conditional neural processes for local climate downscaling. \textit{Geoscientific Model Development}, \textbf{15~(1)}, 251--268.

\bibitem[{Wilson et~al.(2015)Wilson, Hu, Salakhutdinov,, and Xing}]{wilson_deep_2015}
Wilson, A.~G., Z.~Hu, R.~Salakhutdinov, and E.~P. Xing, 2015: Deep {Kernel} {Learning}. arXiv, \urlprefix\url{http://arxiv.org/abs/1511.02222}, arXiv:1511.02222 [cs, stat], \doi{10.48550/arXiv.1511.02222}.

\bibitem[{Wilson et~al.(2021)Wilson, Borovitskiy, Terenin, Mostowsky,, and Deisenroth}]{wilson_pathwise_2021}
Wilson, J.~T., V.~Borovitskiy, A.~Terenin, P.~Mostowsky, and M.~P. Deisenroth, 2021: Pathwise {Conditioning} of {Gaussian} {Processes}. arXiv, \urlprefix\url{http://arxiv.org/abs/2011.04026}, arXiv:2011.04026 [cs, math, stat], \doi{10.48550/arXiv.2011.04026}.

\bibitem[{Winstral et~al.(2002)Winstral, Elder,, and Davis}]{winstral_spatial_2002}
Winstral, A., K.~Elder, and R.~E. Davis, 2002: Spatial {Snow} {Modeling} of {Wind}-{Redistributed} {Snow} {Using} {Terrain}-{Based} {Parameters}. \textit{Journal of Hydrometeorology}, \textbf{3~(5)}, 524--538, \doi{10.1175/1525-7541(2002)003<0524:SSMOWR>2.0.CO;2}.

\bibitem[{Zanetta et~al.(2023)Zanetta, Nerini, Beucler,, and Liniger}]{zanetta_physics-constrained_2023}
Zanetta, F., D.~Nerini, T.~Beucler, and M.~A. Liniger, 2023: Physics-{Constrained} {Deep} {Learning} {Postprocessing} of {Temperature} and {Humidity}. \textit{Artificial Intelligence for the Earth Systems}, \textbf{2~(4)}, \doi{10.1175/AIES-D-22-0089.1}.

\bibitem[{Zhuang et~al.(2020)Zhuang, Tang, Ding, Tatikonda, Dvornek, Papademetris,, and Duncan}]{zhuang_adabelief_2020}
Zhuang, J., T.~Tang, Y.~Ding, S.~Tatikonda, N.~Dvornek, X.~Papademetris, and J.~S. Duncan, 2020: {AdaBelief} {Optimizer}: {Adapting} {Stepsizes} by the {Belief} in {Observed} {Gradients}. arXiv, \urlprefix\url{http://arxiv.org/abs/2010.07468}, arXiv:2010.07468 [cs, stat], \doi{10.48550/arXiv.2010.07468}.

\end{thebibliography}

\end{document}